\documentclass[seceq]{ptptex}

\usepackage{graphicx}
\usepackage{wrapft}

\def\lesim{\m@thcombine<\sim}
\def\gesim{\m@thcombine>\sim}
\def\lessgtr{\m@thcombine<>}
\def\gtrless{\m@thcombine><}

\newcommand{\bra}[1]{\left\langle #1 \right|}
\newcommand{\ket}[1]{\left| #1 \right\rangle}
\newcommand{\Hhat}{\hat{H}}
\newcommand{\HhatMq}{\hat{H}_M(q)}
\newcommand{\hhat}{\hat{h}}
\newcommand{\hhatq}{\hat{h}(q)}
\newcommand{\hhatMq}{\hat{h}_M(q)}
\newcommand{\Fhat}{\hat{F}}
\newcommand{\Fhatd}{\hat{F}^\dagger}
\newcommand{\Fhatp}{\hat{F}^{(+)}}
\newcommand{\Fhatm}{\hat{F}^{(-)}}
\newcommand{\Hc}{{\cal H}}
\newcommand{\Nhat}{\hat{N}}

\newcommand{\Qhat}{\hat{Q}}
\newcommand{\Rhat}{\hat{R}}
\newcommand{\Phat}{\hat{P}}
\newcommand{\That}{\hat{\Theta}}

\newcommand{\phiqppn}{\phi(q,p,\varphi,N)}
\newcommand{\phiqpn}{\phi(q,p,N)}

\newcommand{\qpN}{(q,p,N)}
\newcommand{\phiq}{\phi(q)}

\newcommand{\del}{\partial}

\newcommand{\beq}{\begin{equation}}
\newcommand{\beqa}{\begin{eqnarray}}
\newcommand{\eeq}{\end{equation}}
\newcommand{\eeqa}{\end{eqnarray}}

\def\vect#1{\mbox{\boldmath $#1$}}

\title{       
Application of the Adiabatic Selfconsistent-Collective-Coordinate Method\\
to a Solvable Model of Prolate-Oblate Shape Coexistence}

\author{Masato  \textsc{Kobayasi}$^{1}$,Takashi
\textsc{Nakatsukasa}$^{2}$, Masayuki \textsc{ Matsuo}$^{3}$ \\
and Kenichi \textsc{Matsuyanagi}$^1$
}

\inst{
$^1$ Department of Physics, Graduate School of Science,\\
 Kyoto University, Kitashirakawa, Kyoto 606-8502\\
$^2$ Physics Department, Tohoku University,Sendai 980-8578\\
$^3$ Graduate School of Science and Technology,\\
 Niigata University, Niigata 950-2181
}

\abst{
The adiabatic selfconsistent collective coordinate method is applied to
an exactly solvable multi-$O(4)$ model which simulates nuclear shape 
coexistence phenomena. 
Collective mass and
dynamics of large amplitude collective motions in this model system
are analysed, and it is shown that the method can well describe
the tunneling motions through the barrier between
the prolate and oblate local minima in the collective potential.
Emergence of the doublet pattern is well reproduced.
}

\begin{document}

\maketitle

\section{Introduction}

Microscopic description of large amplitude collective motion in nuclei
is a long-standing fundamental subject of nuclear structure physics
\cite{rf:1,rf:2}.
In spite of the steady development in various theoretical concepts 
and mathematical formulations of them, application of the microscopic 
many-body theory to actual nuclear phenomena still remains as 
a challenging subject$^{3)-30)}$
(see Ref.~\citen{rf:31} for a recent comprehensive review). 
Shape coexistence phenomena are typical examples of the nuclear large 
amplitude collective motion and have been investigated from various points
of view$^{32)-45)}$
For instance, even in typical spherical nuclei 
like Pb and Sn isotopes, excited deformed states have been systematically
observed in low-energy regions\cite{rf:32}, 
and the coexistence of prolate, spherical and oblate shapes have been 
recently reported for $^{186,188}$Pb\cite{rf:38}. 
As another example, we mention a recent discovery of
coexisting two rotational bands in $^{68}$Se, which are
associated with oblate and prolate intrinsic shapes\cite{rf:39}.
These are only a few examples among abundant experimental data.
Clearly, these data strongly calls for further development of the theory
which is able to describe them 
and renews our concepts of nuclear structure.
From the viewpoint of the microscopic mean-field theory,
these phenomena imply 
that different solutions of the Hartree-Fock-Bogoliubov (HFB) equations
(local minima in the deformation energy surface) appear 
in the same energy region and the nucleus undergoes a
large amplitude collective motions connecting these different
equilibrium points. Identities or mixings of these different shapes
are determined by the dynamics of such collective motions. 

On the basis of the time-dependent Hartree-Fock (TDHF) theory,
the selfconsistent collective coordinate (SCC) method was proposed
as a microscopic theory of such large amplitude collective motions
\cite{rf:9}.
It was extended to the case of time-dependent HFB (TDHFB) 
including the pairing correlations\cite{rf:20}, and
has been successfully applied to various kinds of anharmonic vibration 
and high-spin rotations$^{46)-57)}$
In order to apply this method to shape coexistence phenomena, however,
we need to further develop the theory, since the known method of 
solving the basic equations of the SCC method,
called $\eta$-expansion method\cite{rf:9}, assumes a single local minima
whereas several local minima of the potential energy surface 
compete in these phenomena. Quite recently, a new method
of solving the basic equations of the SCC method, 
called adiabatic SCC (ASCC) method, has been proposed\cite{rf:58}. 
This new method uses an expansion in terms of the collective momentum
and do not assume a single local minima, so that it is expected to be
suitable for a description of the shape coexistence phenomena.
The ASCC method may also be regarded as a successor of the
adiabatic TDHF (ATDHF) methods.
It inherits the major achievements of the ATDHF theory
(as reviewed in Ref.~\citen{rf:31})
and, in addition,
enables us to include the pairing correlations selfconsistently removing
the spurious number fluctuation modes.

The major purpose of this paper is to examine the feasibility of
the ASCC method for applications to actual nuclear phenomena.
This is done by applying it to an exactly solvable model called
multi-$O(4)$ model and testing results of the ASCC method against
exact solutions obtained by diagonalizating the Hamiltonian 
in a huge bases. This solvable model may be regarded as a simplified version
of the well-known pairing-plus-quadrupole (P+Q) interaction model
\cite{rf:59,rf:60}. 
Namely, only the $K=0$ component of the quadrupole deformation is 
considered in a schematic manner, and the model
has been widely used as a testing ground for various microscopic 
theories of nuclear collective motion\cite{rf:61,rf:62,rf:63,rf:64}.
The multi-$O(4)$ model possesses a symmetry, analogous to the ordinary
parity quantum number, with respect to the sign change of the
``quadrupole'' deformation. Accordingly, it can be utilized 
as a simple model of many-body systems possessing
double well structure where large amplitude tunneling motions take place
through the barrier between the two degenerate local minima of the potential
(which correspond to the prolate and oblate shapes).
Because of the special symmetry of the model, 
the ``prolate'' and ``oblate'' shapes mix completely.
Of course, in contrast to the ordinary parity, 
such an exact prolate-oblate symmetry does not occur in reality, 
and in this sense this solvable model is somewhat unrealistic. 
Nevertheless, using this model, we can make a severe test
of the theory by examining its ability to describe the ``parity doublet''
pattern.

In this paper, we would like to focus our attention on 
the collective dynamics and the collective mass of large
amplitude collective motions.
Needless to say, the barrier penetration
depends on the collective mass in a quite sensitive manner.
A similar investigation on the collective dynamics of this model
was done in Ref.~\citen{rf:27},
in which the number fluctuation's degree of freedom was explicitly
removed from the model space.
In the present approach,
the spurious number fluctuation modes are automatically
decoupled from the physical modes within a selfconsistent framework 
of the TDHFB theory.  
This will be a great advantage when the method is applied to realistic
nuclear problems.

This paper is arranged as follows: 
In $\S$ 2, the basic equations of the ASCC method are recapitulated. 
In $\S$ 3, a brief account of the multi-$O(4)$ model is given.
In $\S$ 4, we apply the ASCC method to the multi-$O(4)$ model and
derive explicit expressions necessary for numerical calculations.
In $ \S$ 5, we present results of numerical analysis, and
conclusions are given in $ \S$ 6.

\section{Basic equations of the ASCC method}

In this section, 
instead of recapitulating the  general outline of the SCC method, 
we summarize the adiabatic version of them
formulated in Ref.~\citen{rf:58}.
We assume that large-amplitude collective motions are described 
by a set of the TDHFB state vectors $\ket\phiqppn$
that are parametrized by a single collective coordinate $q$,
the collective momentum $p$ conjugate to $q$,
the particle number $N$ and the gauge angle $\varphi$ conjugate to $N$.
The time evolution of $\ket\phiqppn$ is determined by the
time-dependent variational principle

\begin{equation}
\delta\bra\phiqppn i\frac{\partial}{\partial t} 
-{\hat H} \ket\phiqppn=0.
\label{inv}
\end{equation}

\noindent
As discussed in Ref.~\citen{rf:58}, we can set
\begin{equation}\label{rotating}
\ket{\phiqppn}=e^{-i \varphi \Nhat}\ket{\phiqpn}.
\end{equation}
Then, since the Hamiltonian $\Hhat$ commutes with the number operator
$\Nhat$, the gauge angle $\varphi$ becomes cyclic.
The basic equation of the SCC method consists of the invariant principle
of the TDHFB equations (\ref{inv}) and the canonical variable condition which
can be written as equations for the state vectors 
$\ket{\phi\qpN}$,
\begin{subequations}
\label{cvc2}
\begin{eqnarray}
\bra{\phiqpn}i{\del\over\del q}\ket{\phiqpn} & =p, \\ 
\bra{\phiqpn}{\del\over i\del p}\ket{\phiqpn} & =0,\\ 
\bra{\phiqpn}\Nhat\ket{\phiqpn} & =N, \\ 
\bra{\phiqpn}{\del\over i \del N}\ket{\phiqpn} & =0. 
\end{eqnarray}
\end{subequations}

\noindent
The third equation guarantees that the particle-number expectation 
value is kept constant during the large-amplitude collective motion
described by the collective variables $(q,p)$.
Assuming that the large-amplitude collective motion described by 
the collective variables $(q,p)$ is slow, we now introduce the
adiabatic approximation to the SCC method.
Namely we expand the basic equations with respect to the collective
momentum $p$. Since the particle-number variable $N$
is a momentum variable in the present formulation,
we also expand the basic equations with respect to $n=N-N_0$,
when we consider a system with particle number $N_0$.
We then keep only the lowest order term. The TDHFB state vectors are
thus written as

\begin{equation}
|\phi(q,p,N)\rangle=e^{ip{\hat Q(q)}+in{\hat \Theta(q)}}|\phi(q)\rangle,
\label{ketqpn}
\end{equation}

\noindent
where $\Qhat(q)$ and $\That(q)$ are infinitesimal generators
with respect to $\ket{\phiq}$.
We also define an infinitesimal generator $\Phat(q)$ by

\begin{equation}
e^{-i \delta q {\hat P(q)}}|\phi(q)\rangle=|\phi(q+\delta q)\rangle.
\label{deltaq}
\end{equation}
We insert the TDHFB state vectors (\ref{ketqpn})
into (\ref{inv}) and make an expansion with respect to
$p$ and $n$. Requiring that the time-dependent variational principle 
(the canonical variable condition) be fulfilled up to 
the second (first) order,
we obtain the basic set of equations of the ASCC method to determine the
infinitesimal generators  $\Qhat(q)$ and $\Phat(q)$ as follows:\\

\underline{Canonical variable conditions}
\begin{subequations}
 \label{cvcqp}
\begin{eqnarray}
&& \bra{\phiq}[\Qhat(q),\Phat(q)]\ket{\phiq} = i, \\
&& \bra{\phiq}[\That(q),\Nhat]\ket{\phiq} = i, 
\end{eqnarray}

\end{subequations}

\noindent
the other expectation values of commutators among 
$\{\Qhat(q),\Phat(q),\That(q),\Nhat \}$ being zero.\\

\underline{HFB equation in the moving frame}\\
\begin{equation} 
\delta\bra{\phiq}\HhatMq \ket{\phiq} = 0,   
\label{eqcsmf}
\end{equation}
\noindent
where the $\HhatMq$ is the Hamiltonian in the moving frame defined by
\begin{equation}
\HhatMq=\hat{H}-\lambda(q)\hat{N}-\frac{\partial V}{\partial q}\hat{Q}(q).
\end{equation}

\underline{Local harmonic equations}\\
\begin{subequations}
\begin{equation} 
\delta\bra{\phiq}[\HhatMq, \Qhat(q) ] - {1\over i} B(q) \Phat(q)
\ket{\phiq} = 0,   
\label{eqcshq}
\end{equation}

\begin{equation}
\delta\bra{\phiq} [\HhatMq, {1\over i}\Phat(q)] -C(q)\Qhat(q) \\\\
-{1 \over 2B(q)}[[\HhatMq, (\Hhat - \lambda(q)\Nhat)_{A}], \Qhat(q)]
-{\del \lambda \over \del q}\Nhat
\ket{\phiq} = 0,  
\label{eqcshp}
\end{equation}
\end{subequations}

\noindent
where the local stiffness $C(q)$ is defined by
\begin{equation}
C(q) = {\del^2 V \over \del q^2} 
+ {1\over 2B(q)}{\del B\over \del q}{\del V \over \del q} ,
\end{equation}
\noindent
and $(\Hhat - \lambda\Nhat)_{A}$ indicates
the $a^\dagger a^\dagger$ and $aa$ parts of 
($\Hhat - \lambda\Nhat$) containing two-quasiparticle creation and
annihilation operators.
The collective potential $V(q)$, the inverse mass $B(q)$,
and the chemical potential $\lambda(q)$ are defined below.
Equations (\ref{eqcshq}) and (\ref{eqcshp}) are linear equations
with respect to the one-body operators $\Qhat(q)$ and $\Phat(q)$.
They have essentially the same structure as
the standard RPA equations 
except for the last two terms in Eq.~(\ref{eqcshp}),
which arise from the curvature term (derivative of the generator)
and the particle-number constraint, respectively.
The infinitesimal
generators $\Qhat(q)$ and $\Phat(q)$ are thus closely related to
the harmonic normal modes locally defined for $\ket{\phiq}$ and
the moving frame Hamiltonian $\HhatMq$.
The collective subspace defined by these equations
will be uniquely determined 
once a suitable boundary condition is specified. 

The collective Hamiltonian is given by
\begin{subequations}
\begin{eqnarray}
 \Hc(q,p,N) &\equiv& \bra{\phiqpn}\Hhat\ket{\phiqpn} \\  
\label{hcol}
            &=& V(q) + {1\over 2} B(q) p^2 + \lambda(q)n,
\end{eqnarray}
\end{subequations}
\noindent
up to the second order in $p$ and the first order in $n$, where
\begin{eqnarray}         
\label{pot}
 V(q) &=& \Hc\qpN|_{p=0,N=N_0} = \bra{\phiq}\Hhat\ket{\phiq}, \\
 B(q) &=&{1\over2}{\del^2\Hc\qpN\over\del p^2}|_{p=0,N=N_0} = 
           -\bra{\phiq}[[\Hhat,\Qhat(q)],\Qhat(q)]\ket{\phiq}, \\
\label{mass}
 \lambda(q) &=&{\del\Hc\qpN\over\del N}|_{p=0,N=N_0}
                 = \bra{\phiq}[\Hhat,i\That(q)]\ket{\phiq}. 
\label{hcolexpand}
\end{eqnarray}
For the system with $N=N_0$ particles, we can put $n=0$.

\section{Multi-$O(4)$ model}

The multi-$O(4)$ model may be regarded 
as a simplified version of
the conventional P+Q interaction model\cite{rf:59,rf:60},
where only the $K=0$ component 
of the quadrupole deformation is considered in a schematic manner.
It has been used for schematic analysis of
anharmonic vibrations in transitional nuclei and of various kinds of
large-amplitude collective motion\cite{rf:61,rf:62,rf:63,rf:64}.

We define bilinear fermion operators for each $j$-shell,
\begin{subequations}
\begin{eqnarray}
A_j^\dag &=& \sum_{m>0} c_{jm}^{\dag} c_{j-m}^{\dag},~~
B_j^\dag = \sum_{m>0} {\sigma_{jm}}c_{jm}^{\dag} c_{j-m}^{\dag},~\\
\hat{N}_j &=& \sum_m c_{jm}^{\dag} c_{jm},~~~~~
\hat{D}_j = \sum_m {\sigma_{jm}} c_{jm}^{\dag} c_{jm},
\end{eqnarray}
\end{subequations}
\noindent
with
\begin{equation}
{\sigma_{jm}} =
\left \{
\begin{array}{cc}
 1 &  |m| < \Omega_{j}/2 \\
-1 &  |m|  >   \Omega_{j}/2. 
\end{array}
\right.
\end{equation}
The sign of $\sigma_{jm}$ is chosen so as to simulate the behavior of
the quadrupole matrix elements $\bra {jm} r^2Y_{20} \ket{jm}$,
and we assume that the pair multiplicity $\Omega_j=j+\frac{1}{2}$
is an even integer.
The set of operators $\{A_j^\dag,A_j,B_j^\dag,B_j,\hat{N}_j,\hat{D}_j\}$
form a Lie algebra of $O(4)$.
We then define their extensions to the multi $j$-shell case,

\begin{equation}
A^\dag = \sum_j A^\dag_j,~
B^\dag = \sum_j B^\dag_j,~
\hat{N} = \sum_j \hat{N}_j,~
\hat{D} = \sum_j d_j \hat{D}_j,~
\end{equation}
the coefficients $d_j$ in $\hat{D}$ simulating the magnitudes of 
the reduced quadrupole matrix elements of the $j$-shells,
and introduce a model Hamiltonian
\begin{subequations}
\label{Ho4}
\begin{eqnarray}
\hat{H} &=& \hat{h}_0 - \frac{1}{2} G (A^{\dag}A + AA^{\dag})
                    - \frac{1}{2} \chi {\hat D}^{2}, \\ 
\hat{h}_0 &=& \sum_j e^0_j{\hat N}_j,
\end{eqnarray}
\end{subequations}
where $e^0_j$ denote single-particle energies of the $j$-shells,
and $G$ and $\chi$ represent the strengths of the pairing and
the ``quadrupole'' interactions, respectively.
Note that this Hamiltonian is invariant with respect to 
the conversion of single-particle states 
$(j,\pm|m|) \leftrightarrow (j, \pm(\Omega_j-|m|))$.  
Thus, eigenstates can be
classified according to the ``parity'' quantum number 
associated with this symmetry.

If we make different combinations of these operators as
\begin{subequations}
\begin{eqnarray}
K_{j+}&=&\frac{1}{2}(A_j^\dag + B_j^\dag),~~~~~~~~~
L_{j+}=\frac{1}{2}(A_j^\dag - B_j^\dag),~~\\
K_{j-}&=&\frac{1}{2}(A_j + B_j),~~~~~~~~~
L_{j-}=\frac{1}{2}(A_j - B_j),~~\\
K_{j0}&=&\frac{1}{2}(\hat{N}_j + \hat{D}_j -\Omega_j),~~
L_{j0}=\frac{1}{2}(\hat{N}_j - \hat{D}_j -\Omega_j),
\end{eqnarray}
\end{subequations}
the sets $(K_{j+}, K_{j-},K_{j0})$ and $(L_{j+}, L_{j-},L_{j0})$
separately form $SU(2)$ algebras, and they commute with each other.
Namely, the multi-$O(4)$ model is equivalent to 
the multi $SU(2) \otimes SU(2)$ model.
Thus, we can diagonalize the Hamiltonian (\ref{Ho4})
in a basis set
\begin{equation}
\prod_j | n_{Kj},n_{Lj} \rangle = 
\prod_j (K_{j+})^{n_{Kj}} (L_{j+})^{n_{Lj}} | 0 \rangle,
\label{base}
\end{equation}
to get the exact eigenvalues and eigen-vectors,
where $n_{Kj}$ and $n_{Lj}$ respectively indicate numbers of the $K$
and $L$ pairs in the $j$-shell.
They satisfy $0 \le n_{Kj}, n_{Lj} \le 
\Omega_j/2$ and $\sum_j (n_{Kj} + n_{Lj}) = N_0 /2$.
We note that, 
in the special case that single-particle levels $e^0_j$ are equidistant,
all $d_j$ are equal, and all $\Omega_j=2$,
this model reduces to the one used in Ref.~\citen{rf:65} which studied
collective mass in finite superconducting systems.

\section{Application of the ASCC method to the multi-$O(4)$ model}

\subsection{Quasiparticle representation}

We are now in a position to apply the ASCC method to the multi-$O(4)$ model.
For separable residual interactions such as those in this model,
it is customary to neglect  their Fock terms \cite{rf:59,rf:60}.
We follow this prescription.
Accordingly, in the following, we use the notation HB in place of HFB. 
It is readily seen that the TDHB state vectors $\ket{\phiq}$ 
in the multi-$O(4)$ model can be written in the BCS form,
\begin{equation}
\ket\phiq = \exp\{\sum_i \theta_i (q)(A_i^\dag - A_i)\} |0\rangle,
\label{BCS}
\end{equation}
where $\theta_i (q)$ are related with 
the coefficients $u_i$ and $v_i$ of the Bogoliubov transformation 
to the quasiparticle operators $a_i^{\dagger}$ and $a_i$,   
\begin{equation}           
\left(
\begin{array}{c}
      {a_{i}}^{\dagger}\\
      {a_{-i}}
\end{array}
\right)
 \equiv
\left(
\begin{array}{cc}
           u_{i} & -v_{i}  \\
           v_{i} & u_{i} 
\end{array}
\right)
\left(
\begin{array}{c}
      {c_{i}}^{\dagger}\\
      {c_{-i}}
\end{array}
\right),
\label{bog}
\end{equation}
as $u_i=\sin \theta_i$ and $v_i=\cos \theta_i$.
Here, $i \equiv (j,m)$,~$-i \equiv (j,-m)$, 
and $\sum_i$ denotes a sum over levels with $m > 0$.
We use these conventions hereafter.

The pair operator
$A_{i}^{\dag} \equiv c_{i}^{\dag}c_{-i}^{\dag}$
and the number operator
${\hat N_{i}} \equiv c^{\dag}_{i}c_{i}+c^{\dag}_{-i}c_{-i}$
for the degenerate single-particle levels $(i, -i)$
are written in terms of the {\it quasiparticle}-pair,
${\vect{A}}_{i}^{\dag} \equiv a^{\dag}_{i}a^{\dag}_{-i}$
and {\it quasiparticle}-number operator,
${\hat{\vect{N}}}_{i} \equiv a^{\dag}_{i}a_{i}+a^{\dag}_{-i}a_{-i}$,
as
\begin{subequations}
\begin{eqnarray}
 A_{i}^\dag &=& u_{i}v_{i}+u_{i}^{2}{\vect{A}}_{i}^{\dag}
       -v_{i}^{2}{\vect{A}}_{i}
       -u_{i}v_{i}{\hat {\vect{N}}}_{i}, \\
 {\hat N_{i}} &=& 2v_{i}^{2}+2u_{i}v_{i}({\vect{A}}_{i}^{\dag}+{\vect{A}}_{i})
       +(u_{i}^{2}-v_{i}^{2}){\hat {\vect{N}}}_{i}.
\end{eqnarray}
\end{subequations}
The quasiparticle operators, ${\vect{A}}_{i}^{\dag},~{\vect{A}}_{i}$,
and ${\hat{\vect{N}}}_{i}$, satisfy the commutation relations 
\begin{subequations}
\begin{eqnarray}
\left[{\vect{A}}_{i},{\vect{A}}_{i'}^{\dag}\right]
     &=& \delta_{ii'}(1-{\hat {\vect{N}}}_{i}), \\  
\left[{\hat{\vect{N}}}_{i},{\vect{A}}_{i'}^{\dag}\right]
     &=& 2\delta_{ii'} {\vect{A}}_{i'}^{\dag}. 
\end{eqnarray}
\end{subequations}
We define the deformation $D(q)$ and 
the pairing gap $\Delta(q)$
for  the TDHB state $\ket{\phiq}$ as
\begin{equation}
D(q) = \langle\phi(q)|{\hat D}|\phi(q)\rangle \\
     =  2\sum_{i} d_{i} \sigma_{i}v_{i}^{2},
\label{defo}
\end{equation}
\begin{equation}
\Delta(q)= G \bra{\phiq} A^\dag \ket{\phiq} = G\sum_{i} u_{i}v_{i}. 
\label{delta}
\end{equation}

\subsection{Quasiparticle RPA at local minima}

We start from the standard procedure for describing
small-amplitude vibrations around the local minima of the collective
potential. Namely, we apply the quasiparticle
RPA about the HB equilibrium points.
Since the ASCC is equivalent to the HB+RPA at equilibrium states,
the quasiparticle RPA modes provide
the boundary condition for solving the local harmonic equations of the
ASCC method. 

For the equilibrium HB state $\ket{\phi_0}$, the HB equation is given,
as usual, by
\begin{equation}           
\left(
\begin{array}{cc}
      e_{i}-\lambda_0 & \Delta_0 \\
      \Delta_0 & -e_{i}+\lambda_0  
\end{array}
\right)
\left(
\begin{array}{c}
           u_{i} \\
           v_{i}
\end{array}
\right)
= E_{i}
\left(
\begin{array}{c}
           u_{i} \\
           v_{i}
\end{array}
\right),
\label{HB}
\end{equation}
where $\Delta_0$ and $\lambda_0$ denote
the pairing gap and the chemical potential, and
\begin{equation}
e_i=e^0_i - \chi d_i \sigma_i D_0 
\end{equation}
 are single-particle energies at the equilibrium deformation $D_0$.
The quasiparticle energy and the particle number are
written as
$E_i=\sqrt{(e_i-\lambda_0)^{2}+\Delta_0^2}$ 
and $N_0=2\sum_iv_i^2$.

Writing the RPA normal coordinates and momenta as
\begin{subequations}
\begin{eqnarray}
{\Qhat_n} &=& \sum_i Q_i^n({\vect A_i^{\dag}}+{\vect A_i}), \\
{\Phat_n} &=& i \sum_i P_i^n({\vect A_i^{\dag}}-{\vect A_i}),
\end{eqnarray}
\end{subequations}
we can easily solve the quasiparticle RPA equations,
\begin{subequations}
\begin{equation} 
\delta\bra{\phi_0}[\Hhat-\lambda_0\Nhat, \Qhat_n ] - {1\over i} B_n \Phat_n
\ket{\phi_0} = 0,   
\end{equation}
\begin{equation}
\delta\bra{\phi_0} [\Hhat-\lambda_0\Nhat, {1\over i}\Phat_n] -C_n\Qhat_n 
\ket{\phi_0} = 0,  
\end{equation}
\end{subequations}
where $B_n$ and $C_n$ denote the inverse mass and the stiffness
for the $n$-th RPA solution.
Note that the local harmonic equations (\ref{eqcshq}) and (\ref{eqcshp})
reduce to the RPA equations at the HB equilibrium points,
since the third and fourth terms in Eq.~(\ref{eqcshp}) vanish there.
The RPA dispersion equation 
determining the frequencies $\omega_n=\sqrt{B_nC_n}$ 
is given by
\begin{equation}
 \det(S_{kk'})=0
\end{equation}
where 
$S$ is a $3 \times 3$ matrix composed of
\begin{subequations}
\begin{eqnarray}
S_{11}&=&
\sum_{i}(f_{i}^{2}/(2E_{i}-\omega)+g^{2}_{i}/(2E_{i}+\omega))-1,  \\
S_{12}&=&
\sum_{i}(f_{i}g_{i}/(2E_{i}-\omega)+g_{i}f_{i}/(2E_{i}+\omega)),  \\
S_{13}&=&
\sum_{i}(f_{i}h_{i}/(2E_{i}-\omega)+g_{i}h_{i}/(2E_{i}+\omega)),  \\
S_{22}&=&
\sum_{i}(g^{2}_{i}/(2E_{i}-\omega)+f^{2}_{i}/(2E_{i}+\omega))-1,  \\
S_{23}&=&
\sum_{i}(g_{i}h_{i}/(2E_{i}-\omega)+f_{i}h_{i}/(2E_{i}+\omega)),  \\
S_{33}&=&
\sum_{i}(h^{2}_{i}/(2E_{i}-\omega)+h^{2}_{i}/(2E_{i}+\omega))-1, 
\end{eqnarray}
\end{subequations}
and $S_{k'k}=S_{kk'}$, with the notations
$f_{i}=\sqrt{G}u^{2}_{i},~g_{i}=\sqrt{G}v^{2}_{i}$, and
$h_{i}=2\sqrt{\chi}d_{i}\sigma_{i}u_{i}v_{i}$.
If $\chi=0$, the above dispersion equation reduces to
\begin{eqnarray}        
   \omega^{2}\{(\omega^{2}-4\Delta^{2})
   (\sum_{i}\frac{1}{2E_{i}((2E_{i})^{2}-\omega^{2})} )^{2}
  -(\sum_{i}\frac{u_{i}^{2}-v_{i}^{2}}
     {2E_{i}((2E_{i})^{2}-\omega^{2})} )^{2} \}  
   = 0,
\label{RPAdspkakikae}
\end{eqnarray}
which involves two well-known quasiparticle RPA normal modes;
the pairing vibration $(\omega \simeq 2\Delta)$
and the pairing rotation $(\omega=0)$.
On the other hand, 
if we neglect the residual pairing interactions, 
it reduces to
\begin{equation}
4\sum_{i} \frac{2E_i d_i^2u_i^2 v_i^2}{(2E_i)^{2}-\omega^{2}} 
= \frac{1}{2\chi},
\label{QQrpa}
\end{equation}
which involves a normal mode analogous to the $\beta$ vibrations
in deformed nuclei.
We note that these three kinds of normal modes are decoupled
at the spherical point $(D=0)$ where
$\bra\phiq \left[ \hat{D}, A^\dag + A \right] \ket\phiq = 0$.

\subsection{HB and local harmonic equations in the moving frame}

In order to find a collective subspace in the TDHB space,
we need to solve the RPA-like equations in the moving frame;
the local harmonic equations.
These equations determine generators of the collective space,
$\hat{Q}(q)$ and $\hat{P}(q)$.
Now let us summarize the local harmonic equations
for the multi-$O(4)$ model.
With notations $\Fhat_{s=1} \equiv A$,  $\Fhat_{s=2} \equiv \hat{D}$, 
and 
\begin{equation}
\Fhat_s^{(\pm)} \equiv (\Fhat_s \pm \Fhatd_s)/2 = \pm \Fhat_s^{(\pm)\dagger},
\end{equation}
the multi-$O(4)$ Hamiltonian is written as
\begin{equation}
\hat{H} = \hhat_0
  - \frac{1}{2}\sum_{s=1,2}\kappa_{s}\hat{F}_{s}^{(+)}\hat{F}_{s}^{(+)}
  + \frac{1}{2}\sum_{s=1,2}\kappa_{s}\hat{F}_{s}^{(-)}\hat{F}_{s}^{(-)},
\end{equation}
where the suffices $s=1$ and 2 indicate the pairing and the ``quadrupole''
parts, respectively, and $\kappa_{1}=2G$ and $\kappa_{2}=\chi$.
The equation of motion for the time-dependent mean-field state 
vector $\ket{\phi(q,p)}$ is written as
\begin{equation}
 \delta\bra{\phi(q,p)}i{ \del \over \del t} - 
   \hhat\ket{\phi(q,p)}=0, 
\end{equation}
with the selfconsistent mean-field Hamiltonian
\begin{equation}  
 \hhat = \hhat_0 
        - \sum_s\kappa_s \Fhatp_s \bra{\phi(q,p)}\Fhatp_s\ket{\phi(q,p)}
        + \sum_s\kappa_s \Fhatm_s \bra{\phi(q,p)}\Fhatm_s\ket{\phi(q,p)}.
\end{equation}
The HB equation in the moving frame (\ref{eqcsmf}) and
the local harmonic equations (\ref{eqcshq})-(\ref{eqcshp}) then become

\begin{subequations}
\begin{equation}
\delta\bra{\phiq}\hhatMq \ket{\phiq} = 0,
\end{equation}

\begin{equation} \label{sephq}
\delta\bra{\phiq}[\hhatMq, \Qhat(q) ] - \sum_s f^{(-)}_{Q,s} \Fhatm_s 
- {1\over i} B(q) \Phat(q) 
     \ket{\phiq} = 0,
\end{equation}

\begin{eqnarray} \label{sephp}
\delta\bra{\phiq}[\hhatMq&,& {1\over i}B(q)\Phat(q)] 
    - \sum_s f^{(+)}_{P,s} \Fhatp_s 
    - B(q)C(q)\Qhat(q)
    - \sum_s f^{(+)}_{R,s} \Fhatp_s \nonumber \\
    &-& \sum_s [\Fhatm_s, (\hhat(q)-\lambda(q)\Nhat)_{A}] f^{(-)}_{Q,s} 
    -f_N \Nhat
    \ket{\phiq} =0, 
\end{eqnarray}
\end{subequations}
where $\hhatMq$ is the selfconsistent mean-field Hamiltonian 
in the moving frame defined by
\begin{subequations}
\begin{eqnarray}
&& \hhatMq = \hhatq  - \lambda(q)\Nhat
                     - {\del V\over \del q}\Qhat(q), \\
&&   \hhatq = \hhat_0  
- \sum_s \kappa_s \Fhatp_s \bra{\phiq}\Fhatp_s\ket{\phiq} ,
\end{eqnarray}
\end{subequations}
and  
\begin{subequations}
\begin{eqnarray}
&& f^{(-)}_{Q,s} = - \kappa_s 
              \bra{\phiq}[\Fhatm_s, \Qhat(q)] \ket{\phiq}, \\
&& f^{(+)}_{P,s} = 
   \kappa_s \bra{\phiq}[\Fhatp_s,{1\over i}B(q)\Phat(q)] \ket{\phiq}, \\
&& f^{(+)}_{R,s} = -\frac{1}{2}\kappa_s 
 \bra{\phiq}[[\Fhatp_s,(\hhatq -\lambda(q)\Nhat)_{A}],\Qhat(q)] \ket{\phiq}, \\
&& f_N = B(q){\del \lambda \over\del q}.
\end{eqnarray}
\end{subequations}

We express all operators in the above equations in terms of the
quasiparticle operators $(\vect{A}_i^{\dag}, \vect{A}_i, \hat{\vect{N}_i})$
defined with respect to $\hhatMq$ and $\ket{\phiq}$ as follows: 
\begin{eqnarray}
\hat{h}_{M}(q) &=& \sum_{i}E_{i}{\hat{\vect{N}}}_{i}, \\
\label{Fs}
\hat{F}_{s}^{(+)}&=&
\bra\phiq F_{s}^{(+)}\ket\phiq
+\sum_{i}F_{A,s}^{(+)}(i)({\vect{A}}_{i}^{\dag}+
{\vect{A}}_{i})+\sum_{i}F_{B,s}^{(+)}(i){\hat{\vect{N}}}_{i}, \\
\label{Fhatp}
\hat{F}_{s}^{(-)}&=& 
\sum_{i}F_{A,s}^{(-)}(i)({\vect{A}}_{i}^{\dag}-{\vect{A}}_{i})
\label{Fhatm}
\end{eqnarray}
with 
\begin{subequations}
\begin{eqnarray}  
  F_{A,1}^{(+)}(i)&=&\frac{1}{2}(u_{i}^{2}-v_{i}^{2}),~~~~
  F_{A,2}^{(+)}(i)=2d_{i}\sigma_{i}u_{i}v_{i}, \\ 
  F_{A,1}^{(-)}(i)&=&-\frac{1}{2},~~~~~~~~~~~~~
  F_{A,2}^{(-)}(i)=0, \\ 
  F_{B,1}^{(+)}(i)&=&-u_{i}v_{i},~~~~~~~~~~
  F_{B,2}^{(+)}(i)=d_{i}\sigma_{i}(u_{i}^{2}-v_{i}^{2}). 
\end{eqnarray}
\end{subequations}
Note that all matrix elements are real
so that $\bra\phiq \hat{F_s}^{(-)}\ket\phiq=0$.
For later convenience, we define one-body operators,
\begin{equation}
 \Rhat_s^{(+)}  \equiv 
 [ \Fhat_{B,s}^{(+)}, (\hhat(q)-\lambda(q)\Nhat)_{A}] 
                = 2 \sum_{i}R_s^{(+)}(i) ({\vect A_i^{\dag}} - {\vect A_i}),
\end{equation}
where $\Fhat_{B,s}^{(+)}$ is the last terms of Eq.~(\ref{Fs}) and
\begin{equation}
R_{s}^{(+)}(i)= \{2u_{i}v_{i}(e_i-\chi d_{i}\sigma_{i}-\lambda(q)) - \Delta(u_i^2-v_i^2)\}
                   F_{B,s}^{(+)}(i).
\end{equation}

The infinitesimal generators $\hat{Q}(q)$ and $\hat{P}(q)$ can be witten
as
\begin{subequations}
\begin{eqnarray}
{\hat Q(q)} &=& \sum_{i} Q_{i}(q)({\vect A_{i}^{\dag}}+{\vect A_{i}}), \\
{\hat P(q)} &=& i \sum_{i} P_{i}(q)({\vect A_{i}^{\dag}}-{\vect A_{i}}).
\end{eqnarray}
\end{subequations}
Equations (\ref{sephq}) and (\ref{sephp}) are then reduced to 
linear equations for the matrix elements
$Q_{i}(q)$  and $P_{i}(q)$ of the infinitesimal generators
$\Qhat(q)$  and $\Phat(q)$. They are easily solved to give the expression
\begin{subequations}
\begin{eqnarray}
Q_{i}(q) &=& \frac{2E_{i}}{(2E_{i})^{2}-\omega^2}
\sum_{s} F_{A,s}^{(-)}(i)f_{Q,s}^{(-)}
\nonumber \\
&&+\frac{1}{(2E_{i})^{2}-\omega^2}
\sum_{s}(F_{A,s}^{(+)}(i)f_{PR,s}^{(+)} + N_{i}f_{N}), 
\label{qi} \\
P_{i}(q) &=& \frac{2E_{i}}{(2E_{i})^{2}-\omega^2}
\sum_{s}(F_{A,s}^{(+)}(i)f_{PR,s}^{(+)} + N_{i}f_{N}) \nonumber \\
&&+\frac{\Omega(q)}{(2E_{i})^{2}-\omega^2}
\sum_{s} F_{A,s}^{(-)}(i)f_{Q,s}^{(-)}, 
\label{pi}
\end{eqnarray}
\end{subequations}
where $N_i=2u_i v_i$ and
\begin{equation}
f_{PR,s}^{(+)}=f_{P,s}^{(+)}+f_{R,s}^{(+)}.
\end{equation}
Note that $\omega^2$, representing the square of the 
frequency $\omega(q)=\sqrt{B(q)C(q)}$ of the local harmonic
mode, is not necessarily positive.
The values of $B(q)$ and $C(q)$ depends on the scale of
the collective coordinate $q$, while $\omega(q)$ does not.
In this sense, there remains an ambiguity for determining $q$.
We thus require $B(q)=1$ everywhere on the collective space
to uniquely determine $q$.

The quantities $f_{Q,s}^{(-)},~f_{P,s}^{(+)}$ and $f_{R,s}^{(+)}$
are easily calculated to be
\begin{subequations}
\begin{eqnarray}
f_{Q,s}^{(-)} &=& 2\kappa_{s} \sum_{i} F_{A,s}^{(-)}(i)Q_{i}(q),  
\label{fq} \\
f_{P,s}^{(+)} &=& 2\kappa_{s} \sum_{i} F_{A,s}^{(+)}(i)P_{i}(q), 
\label{fp} \\ 
f_{R,s}^{(+)} &=& 2\kappa_{s} \sum_{i} R_{s}^{(+)}(i)Q_{i}(q).
\label{fr} 
\end{eqnarray} 
\end{subequations}
Inserting  Eqs.~(\ref{qi}) and (\ref{pi}) for $Q_i(q)$ and $P_i(q)$
into the above expressions,
we obtain linear homogeneous equations for 
unknown quantities $f^{(+)}_{PR,s}, f^{(-)}_{Q,s}$  and $f_N$. 
Similarly, the condition of orthogonality to the number operator, 
\begin{equation}
\langle\phi(q)|[\hat{N},\hat{P}(q)]|\phi(q)\rangle 
          = 2i\sum_i N_i P_i(q) = 0,
\end{equation}
gives another equation 
for $f^{(+)}_{PR,s}, f^{(-)}_{Q,s}$  and $f_N$. 
Since $f^{(-)}_{Q,2}=0$, they are summarized 
in a $4 \times 4$ matrix form as follows:
\begin{equation}
\left(
\begin{array}{ccc}
 & & \\
 & S_{kk'}(\omega)& \\
 & & \\
\end{array}
\right) 
\left(
\begin{array}{c}
f^{(-)}_{Q,1} \\
f^{(+)}_{PR,1} \\
f^{(+)}_{PR,2} \\
 f_N \\
\end{array}
\right)  = 0,
\end{equation}
where
\begin{subequations}
\begin{eqnarray}
S_{11} &=& 2G S^{(1)}(F_{A,1}^{(-)},F_{A,1}^{(-)}) -1, \\
S_{12} &=& 2GS^{(2)}(F_{A,1}^{(-)},F_{A,1}^{(+)}),  \\
S_{13} &=& 2GS^{(2)}(F_{A,1}^{(-)},F_{A,2}^{(+)}),  \\
S_{14} &=& 2GS^{(2)}(F_{A,1}^{(-)},N),  \\
S_{21} &=& 2G\{S^{(1)}(R_{1}^{(+)},F_{A,1}^{(-)})
           + \omega^2 S^{(2)}(F_{A,1}^{(+)},F_{A,1}^{(-)})\},   \\
S_{22} &=& 2G\{S^{(1)}(F_{A,1}^{(+)},F_{A,1}^{(+)})
                 + S^{(2)}(R_{1}^{(+)},F_{A,1}^{(+)})\}-1,  \\
S_{23} &=& 2G\{S^{(1)}(F_{A,1}^{(+)},F_{A,2}^{(+)})
                    + S^{(2)}(R_{1}^{(+)},F_{A,2}^{(+)})\},  \\
S_{24} &=& 2G\{S^{(1)}(F_{A,1}^{(+)},N)
                                  + S^{(2)}(R_{1}^{(+)},N)\},  \\
S_{31} &=& 4\chi \{S^{(1)}(R_{2}^{(+)},F_{A,1}^{(-)})
                 + \omega^2 S^{(2)}(F_{A,2}^{(+)},F_{A,1}^{(-)})\},  \\
S_{32} &=& 4\chi \{S^{(1)}(F_{A,2}^{(+)},F_{A,1}^{(+)})
                   + S^{(2)}(R_{2}^{(+)},F_{A,1}^{(+)})\},  \\
S_{33} &=& 4\chi \{S^{(1)}(F_{A,2}^{(+)},F_{A,2}^{(+)})
                  + S^{(2)}(R_{2}^{(+)},F_{A,2}^{(+)})\}-1,  \\
S_{34} &=& 4\chi \{S^{(1)}(F_{A,2}^{(+)},N)
                      + S^{(2)}(R_{2}^{(+)},N)\},  \\
S_{41} &=& \omega^2 S^{(2)}(N,F_{A,1}^{(-)}),  \\
S_{42} &=& S^{(1)}(N,F_{A,1}^{(+)}),  \\
S_{43} &=& S^{(1)}(N,F_{A,2}^{(+)}),  \\
S_{44} &=& S^{(1)}(N,N). 
\end{eqnarray}
\end{subequations}
\noindent
Here, the functions $S^{(1)}(X,Y)$ and $S^{(2)}(X,Y)$ are defined by    
\begin{subequations}
\begin{equation}
 S^{(1)}(X,Y) = \sum_{i} 
              { 2E_i \over (2E_i)^2 - \omega^2} X_{i}Y_{i}, \\
\end{equation}
\begin{equation}
 S^{(2)}(X,Y) = \sum_{i} 
                { 1 \over (2E_i)^2 - \omega^2} X_{i}Y_{i}, \\
\end{equation}
\end{subequations}
with $X_i$ and $Y_i$ denoting ones among the quantities
$F_{A,s}^{(\pm)}(i), R_{s}^{(+)}(i)$ and $N_i$.
The frequency $\omega (q)$ is determined by finding the lowest solution
of the dispersion equation 
\begin{equation}
{\rm det}\{S_{kk'}(\omega)\} = 0,
\end{equation}
i.e., the solution of which $\omega^2$ is minimum.
Normalizations of 
$f^{(+)}_{PR,1}, f^{(+)}_{PR,2}, f^{(-)}_{Q,1}$  and 
$f_N$ are fixed by the canonical variable condition 
\begin{equation}
\langle\phi(q)|[\hat{Q}(q),\hat{P}(q)]|\phi(q)\rangle 
          = 2i\sum_i Q_i(q)P_i(q) = i.
\end{equation}
Note that
$\omega^2$ represents the curvature of the collective potential,
\begin{equation}
\omega^2=\frac{\partial^2 V}{\partial q^2}, 
\label{freq}
\end{equation}
for the choice of coordinate scale such that $B(q)=1$.

\section{Numerical analysis}

In this section, we make numerical analysis of 
the oblate-prolate shape coexistence and 
large-amplitude collective motions in the multi-$O(4)$ model.

\subsection{Procedure of calculation}

We first solve the HB equations and find HB equilibrium points 
which correspond to extrema of the collective
potential $V(q)$ defined by Eq.~(\ref{pot}). 
At these points,
the HB equation in the moving frame, Eq.~(\ref{eqcsmf}), and 
the local harmonic equations, Eqs.~(\ref{eqcshq}) and (\ref{eqcshp}),
coincide with the ordinary HB equation and the quasiparticle RPA
equations, respectively.
Let $\ket{\phi(q_0)}$ be a HB solution, 
which is assumed to be on the collective subspace 
at a particular value of the collective coordinate, $q=q_0$. 
Solving the quasiparticle RPA equation with respect to
$\ket{\phi(q_0)}$, 
we find a collective normal mode, which determines
the infinitesimal generators $\Qhat(q_0)$  and $\Phat(q_0)$. 
In the present analysis, 
we choose the normal mode with the lowest frequency, which is
most collective with respect to to the ``quadrupole'' operator $\hat{D}$.
We then generate the state $\ket{\phi(q_0 + \delta q)}$ 
with an infinitesimal shift of the collective coordinate as

\begin{equation}
 \ket{\phi(q_0 + \delta q)} = 
e^{-i \delta q \Phat(q_0)}\ket{\phi(q_0)} .
\end{equation}
\noindent
Next, we solve the local harmonic equations at $q=q_0+\delta q$ and
determine $\Qhat(q_0+\delta q)$  and $\Phat(q_0+\delta q)$, 
and proceed to $q=q_0 + 2\delta q$.
Repeating this procedure, one can construct a collective subspace.
Thanks to the invariance, 
\begin{equation}
A_i^\dag - A_i = \vect{A}_i^\dag - \vect{A}_i,
\end{equation}
we can rewrite the state vectors as
\begin{eqnarray}
\ket{\phi(q + \delta q)} 
&=& \exp \{\delta q \sum_i P_i (q)(\vect{A}_i^\dag - \vect{A}_i) \} 
\ket{\phi(q)}  \nonumber \\
&=& \exp \{\delta q \sum_i P_i (q)(A_i^\dag - A_i) \}
\ket{\phi(q)}. 
\end{eqnarray}
Combining the above expression with Eq.~(\ref{BCS}), 
we obtain the following simple relations between 
the $u_i, v_i$ coefficients at $q+\delta q$ and those at $q$: 
\begin{equation}
\left(
\begin{array}{cc}
 u_{i}(q+\delta q) \\
 v_{i}(q+\delta q) 
\end{array}
\right)
=
\left(
\begin{array}{cc}
 \cos(P_{i}(q)\delta q) & -\sin(P_{i}(q)\delta q) \\
 \sin(P_{i}(q)\delta q) & \cos(P_{i}(q)\delta q)
\end{array} 
\right)
\left(
\begin{array}{c}
 u_{i}(q) \\
 v_{i}(q) 
\end{array}
\right).
\label{uv}
\end{equation}
\noindent
In the present calculation, we always start from the spherical equilibrium point
($q=0$). This point is an unstable extremum (a saddle point)
on the collective subspace when we choose parameters producing
deformed HB minima (see below).  Note that 
the quasiparticle RPA equations given in subsection 4.1
are still valid whereas the frequency $\omega$ of the eigenmode 
is pure imaginary in this case.

If the collective subspace is exactly decoupled
from the non-collective subspace,
the state vectors $\ket{\phi(q)}$ obtained in this way should satisfy the HB 
equation in the moving frame, Eq.~(\ref{eqcsmf}), simultaneously. 
In general, however, the decoupling conditions may not be exactly satisfied
and we have to resort to some kind of iterative procedure 
to construct a collective subspace which complies with
Eq.~(\ref{eqcsmf}).
One way of evaluating the quality of decoupling 
and accuracy of the numerical calculation is to
examine validity of Eq.~(\ref{eqcsmf}) on the collective space.
For the multi-O(4) model under consideration, 
as we shall see in Fig.~2, this condition is found to be
well satisfied with the use of a step size $\delta q=0.005$ in
Eq.~(\ref{uv}).
Of course, this does not necessarily mean that such a simple algorithm
will always work also for more realistic cases, and we will need more 
investigations concerning numerical techniques 
of solving the set of equations 
(\ref{eqcsmf}), (\ref{eqcshq}), and (\ref{eqcshp}).

The collective Hamiltonian thus obtained,
$\Hc (q,p)=\frac{1}{2}B(q)p^2+V(q)$, is then quantized, 
and the collective Schr\"odinger equation is solved 
to get eigenvalues and transition probabilities.
As we set a scale of the collective coordinate $q$ such that $B(q)=1$,
there is no ambiguity of ordering in the canonical
quantization procedure, following the Pauli's quantization rule.
It is easily confirmed that the collective representation of 
the ``quadrupole'' operator, defined by 
${\cal D}(q) = \bra{\phi(q,p)}{\hat D}\ket{\phi(q,p)}$,
does not depend on $p$, and 
transition matrix elements can be evaluated by
$\int \psi_n(q)^* {\cal D}(q) \psi_{n'}(q) dq$,
where $\psi_n(q)$ denotes
the collective wave function of the $n$-th eigenstate.

\subsection{Parameters}

In the following, we consider a case consisting of three shells
with the spherical single-particle energies,
$e_{j_1}^0=0.0,~e_{j_2}^0=1.0,~e_{j_3}^0=3.5$,
the pair degeneracies,
$\Omega_{j_1}=14,~\Omega_{j_2}=10,~\Omega_{j_3}=4$,
the reduced quadrupole moments,
$d_{j_1}=2,~d_{j_2}=d_{j_3}=1$,
and we distribute 14 pairs ($N_0=28$) in this shell-model space
(we do not distinguish protons and neutrons).
We compare numerical results obtained with different values of
the pairing interaction strength,~$G=0.14,~0.16$, and 0.20,
for a fixed ``quadrupole'' interaction strength $\chi=0.04$. 
These numerical examples are presented merely as representatives
of similar results obtained with other sets of parameters.
Since properties of the single-shell $O(4)$ model
are determined by the ratio $G/\chi$, we obtain similar results varying $\chi$ 
instead of $G$. The only reason why we here vary $G$ fixing $\chi$ is the 
convenience of visualizing the change of the barrier height
(between the prolate and oblate local minima)
while keeping the magnitude of the equilibrium deformation $D_0$ 
almost constant.

\begin{figure}[hhhh]
\begin{center}
{\includegraphics[width=7.5cm]{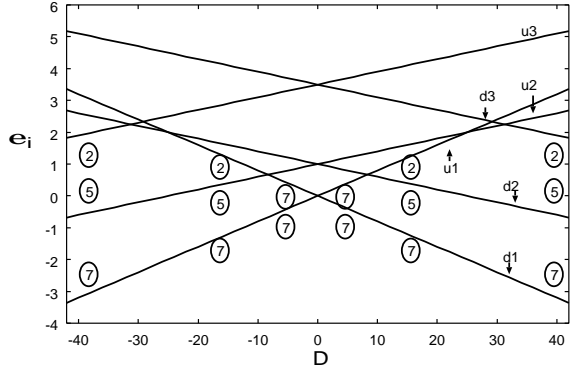}}
\end{center}
\caption{ 
\footnotesize
Deformed single-particle energies, $e_i=e_i^0 - \chi d_i\sigma_i D$, 
plotted as functions of the deformation parameter $D$ 
for a set of parameters:
$e_{j_1}^0=0.0,~e_{j_2}^0=1.0,~e_{j_3}^0=3.5$,
$\Omega_{j_1}=14,~\Omega_{j_2}=10,~\Omega_{j_3}=4$,
and $d_{j_1}=2,~d_{j_2}=d_{j_3}=1$.
Numbers enclosed by circles indicate the numbers of pairs 
occupying individual levels in the lowest energy configuration 
for a given value of $D$ when pairing correlations are absent. 
Near the spherical point ($D=0$), the lowest shell is fully occupied by
14 pairs (28 nucleons).
The level diagram is symmetric with respect to the sign-change of $D$
(prolate-oblate symmetry).
The down(up)-sloping levels (with increasing $D$) are named
``d1,d2, and d3 (u1,u2, and u3)''.
In the prolate side ($D \geq 0$), the down-sloping level originating from
the second shell (d2) crosses the up-sloping level (u1) at $D=8.3$. 
After this first crossing, the lowest-energy configuration involves
5 pairs occupying the second down-sloping level (d2) and 2 pairs remaining
in the first up-sloping level (u1).
At $D=29.2$, the down-sloping level originating from
the third shell (d3) crosses the second up-sloping level (u2).
In the lowest-energy configuration after this crossing,
all the down-sloping levels are fully occupied while
all the up-sloping levels are unoccupied. This configuration
possesses the maximum value of the deformation parameter, $D_{\rm max}=42$.  
The same pattern of level crossings is valid also for the oblate side 
($D \leq 0$). 
Note that this figure is drawn by regarding $D$ as a free parameter,
although $D$ in Eq.~(\ref{defo}) means
an expectation value of the operator $\hat{D}$.
In this sense this figure should be regarded as a kind of Nilsson
diagram.}
\label{fig1}
\end{figure}

\normalsize

Figure 1 illustrates the single-particle energy diagram as
a function of $D$.
Details of the level crossings are explained in the caption
to this figure.
It is certainly possible to discuss the level crossing dynamics 
and simulate the shape coexistence phenomena by means of the multi-$O(4)$
model consisting of only two shells. However, consideration of
the three shells
is found to be more appropriate in order to make properties
in the large deformation region more realistic.
In realistic situations, a number of successive level crossings
take place with increasing deformation.

\subsection{Collective potentials}
\normalsize

We present in Fig.~2 the collective potentials calculated
for three values of the pairing strength $G$.
For $G=0.14$, it exhibits a double well structure, while
it looks like an anharmonic-oscillator potential for $G=0.2$.
The case of $G=0.16$ shows a transitional character between 
these two situations. 
Note that the collective potentials are plotted here as
functions of deformation $D$. The relation
between $D$ and the collective coordinate $q$ will be
explicitly shown later.
The mechanism of appearance and disappearance of the double well
structure is determined by competition between 
the pairing and ``quadrupole'' interactions:
The pairing force favors an equal population of particles over
all magnetic substates in each $j$-shell, whereas the ``quadrupole''
force favors an occupation of down-sloping levels 
creating level crossings and asymmetry in the magnetic substate
population.
In the three $j$-shell case under consideration,
the energy of the degenerate local minima is lower 
than that of the spherical point for small values of $G$.
With increasing pairing correlations, however, the energy of 
the spherical point decreases, so that the barrier hight 
decreases and eventually the barrier itself diminishes.

\begin{figure}[hhhh]
\begin{center}
\begin{tabular}{c}
{\includegraphics[width=7.5cm]{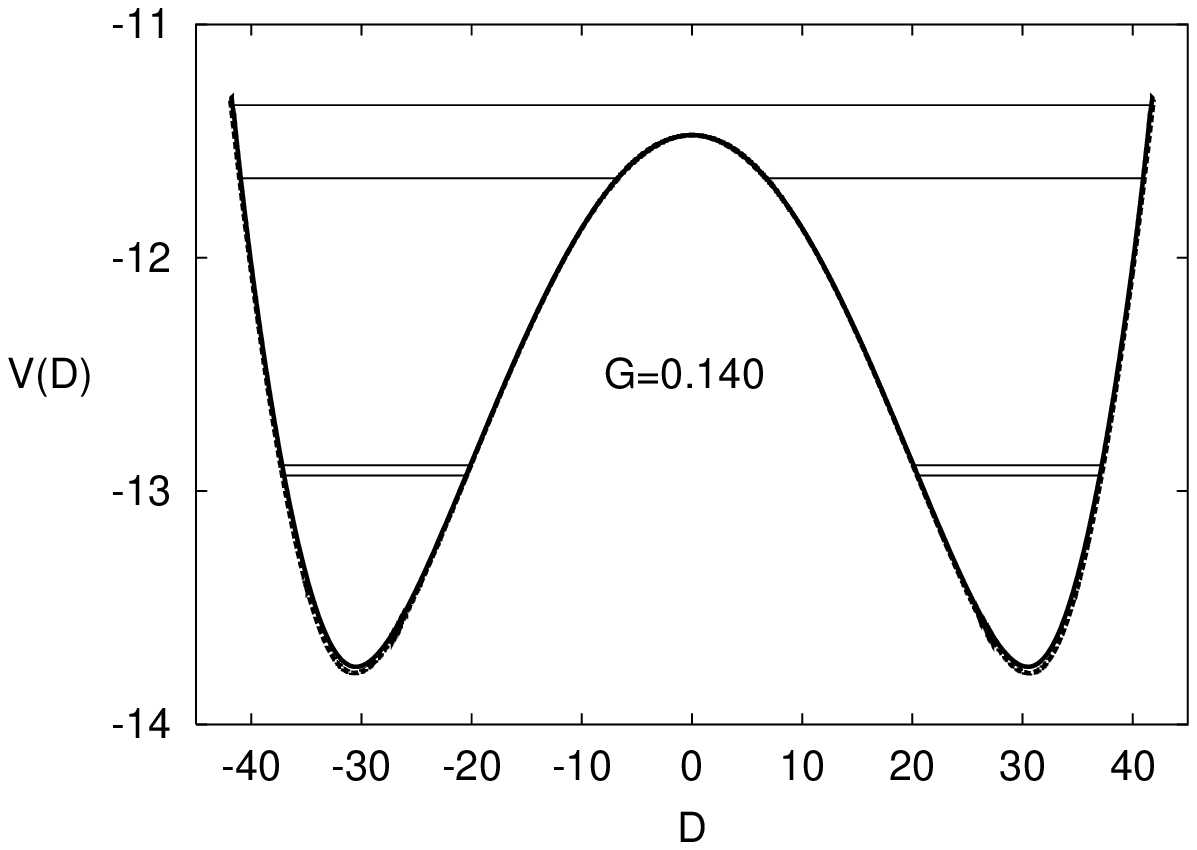}}\\
{\includegraphics[width=7.5cm]{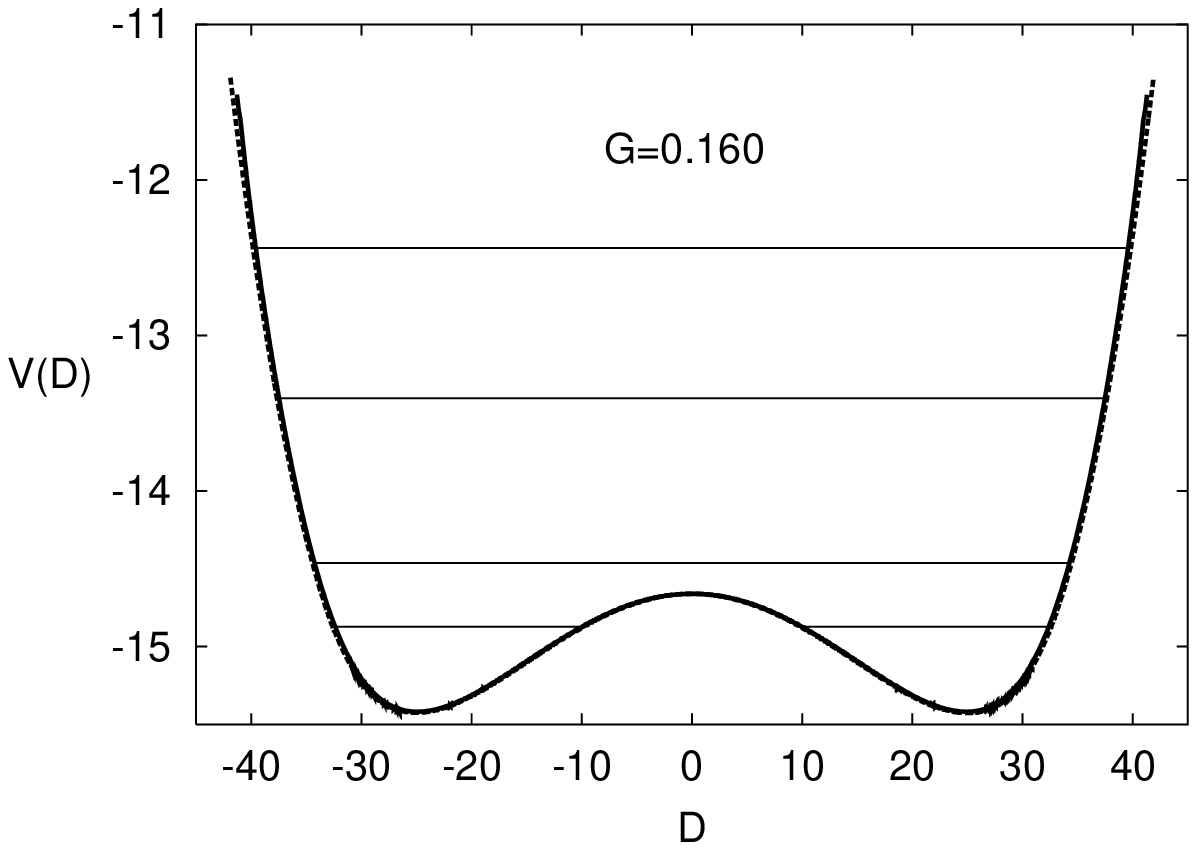}}\\
{\includegraphics[width=7.5cm]{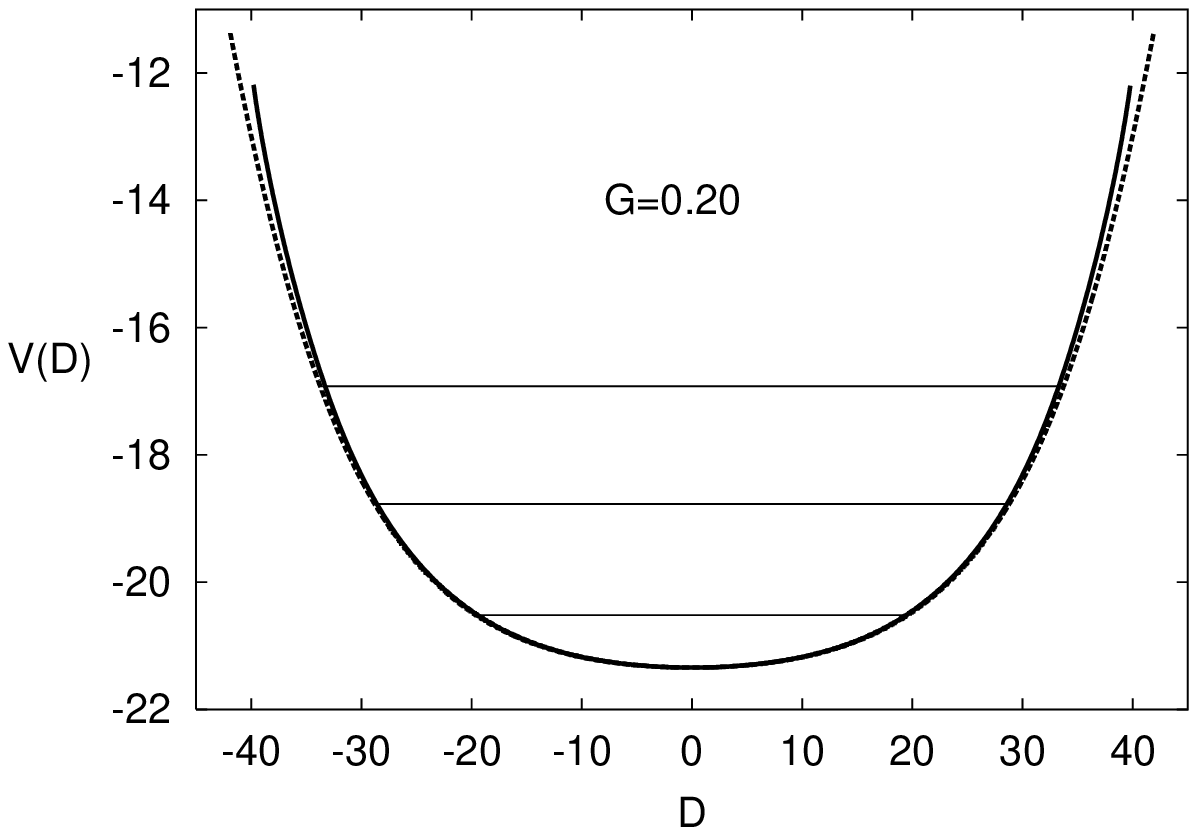}}
\end{tabular}
\end{center}
\caption{
\footnotesize
Collective potentials calculated with the ASCC method 
for $G=0.14$ (top), $G=0.16$ (middle), and $G=0.20$ (bottom)
are plotted by solid lines as functions of deformation $D$. 
Other parameters used in the calculation are listed in the caption 
to Fig.~1. For comparison, collective potentials of the CHB method
are indicated by dotted lines.
Horizontal solid lines indicate eigen-energies of the quantized
collective Hamiltonian.}
\label{fig2}
\end{figure}

\normalsize
In Fig.~2, quantum eigenstates obtained by solving the
collective Shr\"odinger equations are indicated.
We can clearly see a parity-doublet-like pattern
associated with tunneling through the central barrier in the collective
potential curve. Properties of these excitation spectra will be
discussed in the succeeding subsection.
In this figure, we also show the potential energy functions calculated by
means of the conventional constrained HB (CHB) procedure:

\begin{equation}
V_{\rm CHB}(D)=\langle\phi_0^{\rm CHB}(D)|\hat{H}|\phi_0^{\rm CHB}(D)\rangle
= 2 \sum_{i} e^0_{i}v_{i}^{2} -\frac{\Delta^{2}}{G}
                              -\frac{1}{2}\chi D^{2},
\end{equation}
\noindent
where the state vectors $|\phi_0^{\rm CHB}(D) \rangle$ are determined by 
the constrained variational principle 
\begin{equation}
\delta \langle\phi_0^{\rm CHB}(D)|{\hat H} - \lambda {\hat N} 
                                 - \mu {\hat D}|\phi_0^{\rm CHB}(D)\rangle=0,
\label{CHBeq}
\end{equation}
\noindent
with $\mu$ denoting a Lagrange multiplier.
As we see, the collective potentials obtained with the ASCC and 
CHB methods are practically indistinguishable.
In principle, when both the collective path obtained in the ASCC method 
and that in the CHB method go through the same HB local minima, 
the collective potential energies 
at these points should coincide with each other. 
One may notice, however, very small differences between their values
at the HB local minima with $D \ne 0$ for the cases of $G=0.14$ and 0.16.
These discrepancies are due to violation of Eq.~(\ref{eqcsmf})
which accumulates in the numerical calculation starting from $D=0$,
and indicate the amount of error associated with the computational algorithms 
adopted here, as mentioned in subsection 6.1.

In Fig.~3, the pairing gaps $\Delta$ are shown 
as functions of deformation $D$.
They monotonically decrease as $D$ increases, 
and vanish at the maximum deformation,
which is $D_{\rm max}=42$ for the parameters adopted in these calculations.
In Fig.~4, occupation probabilities $v_i^2$ are displayed. 
Note that all down-sloping (up-sloping) levels are fully 
occupied (unoccupied) at $D_{\rm max}$.
Again, the results obtained with the ASCC and CHB methods are practically 
indistinguishable in these figures. 
Thus, in the present multi-$O(4)$ model, the static mean-field properties, 
like the potential energy curves and the pairing gaps,
obtained with the ASCC method,  are very 
similar to those obtained with the conventional CHB method.
This is due to the simplicity of the multi-$O(4)$ model adopted here,
and the collective subspace in general may differ from that of
the CHB method ~\citen{rf:27}.

\begin{figure}[hhhh]
\begin{center}
\begin{tabular}{c}
{\includegraphics[width=7.5cm]{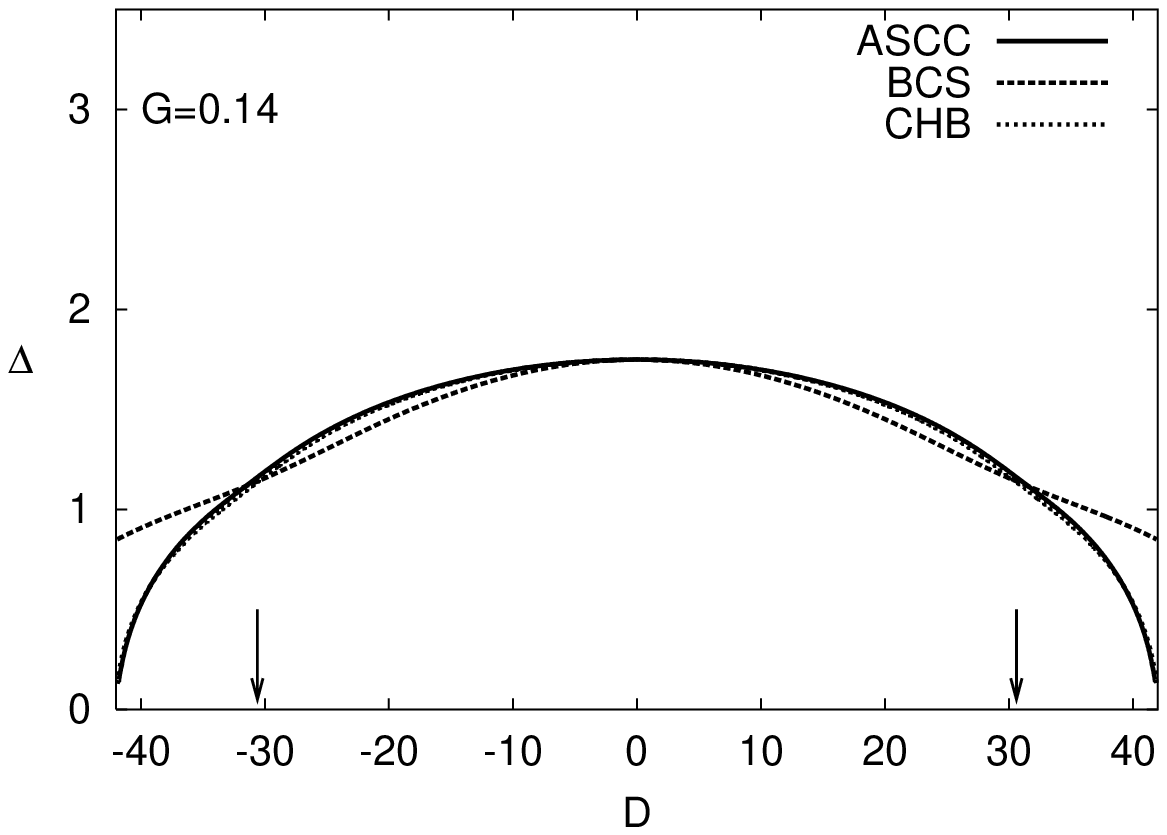}}\\
{\includegraphics[width=7.5cm]{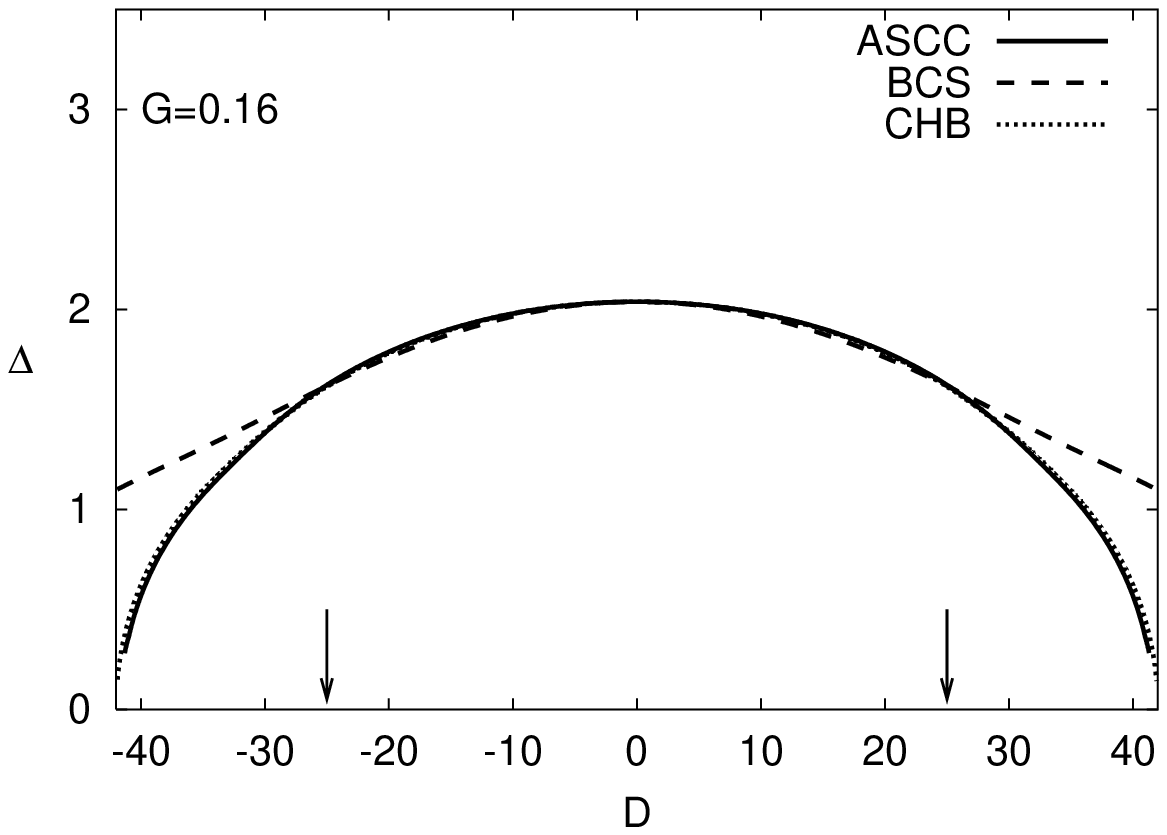}}\\
{\includegraphics[width=7.5cm]{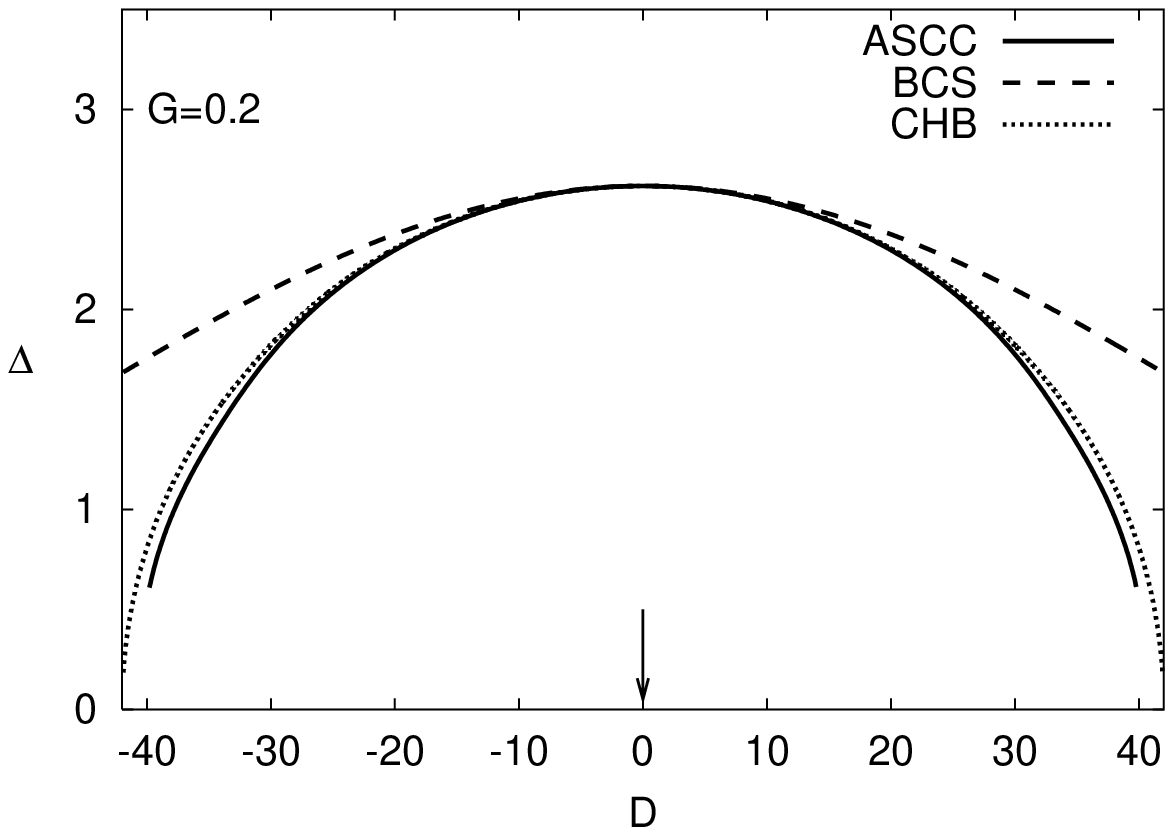}}
\end{tabular}
\end{center}
\caption{ 
\footnotesize 
Pairing gaps $\Delta$ calculated with the ASCC method 
for $G=0.14$ (top), $G=0.16$ (middle), and $G=0.20$ (bottom)
are plotted by solid lines as functions of deformation $D$. 
Other parameters used are the same as in Fig.~2. 
For comparison,  $\Delta$ calculated with the CHB method
and with the BCS approximation are also shown by dotted and dashed lines,
respectively. The solid and dotted lines appear similar to each other.
The equilibrium deformations are indicated by arrows. }
\label{fig3}
\end{figure}

\begin{figure}[hhhh]
\begin{center}
\begin{tabular}{c}
{\includegraphics[width=7.5cm]{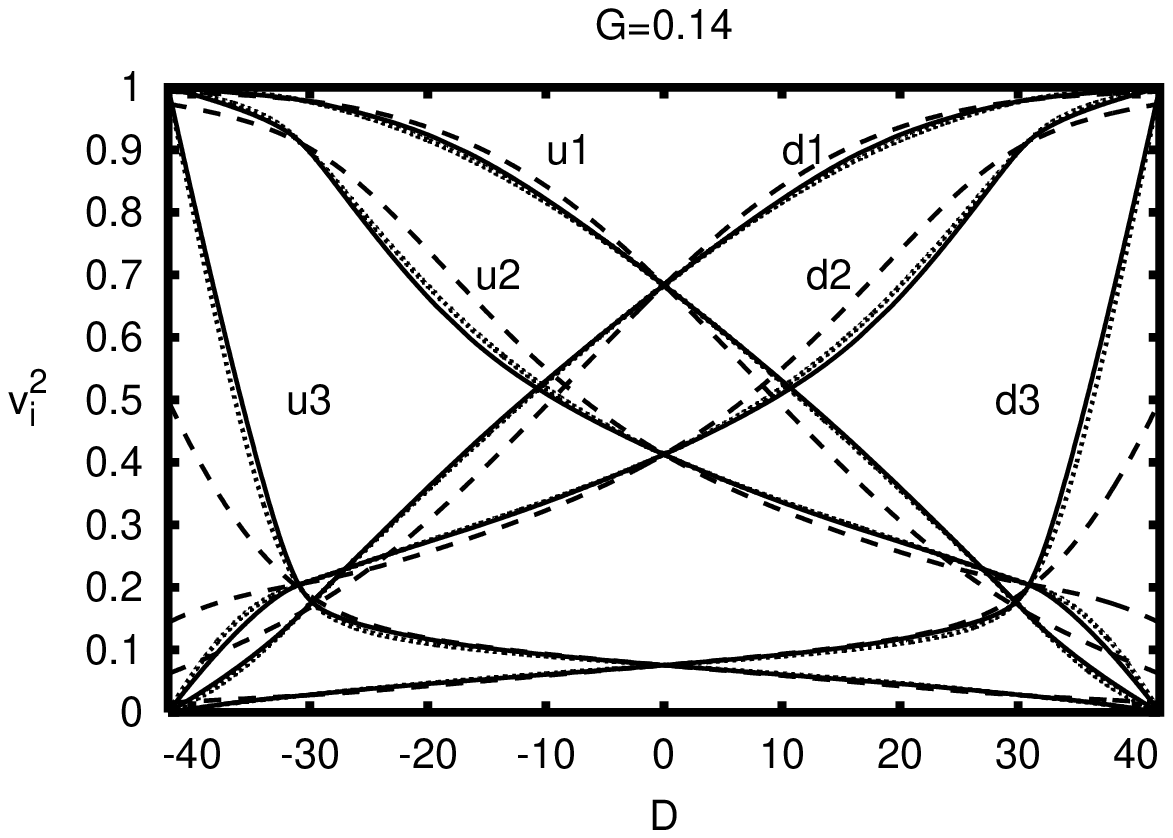}}\\
{\includegraphics[width=7.5cm]{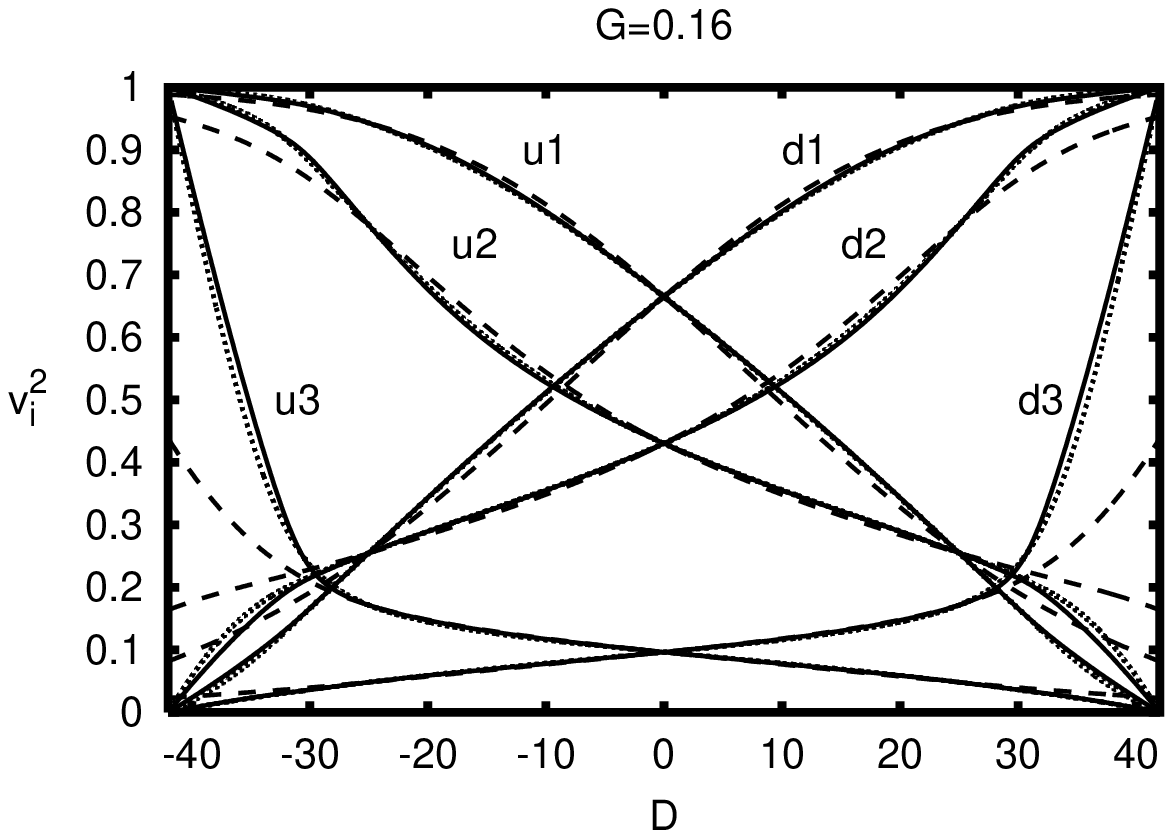}}\\
{\includegraphics[width=7.5cm]{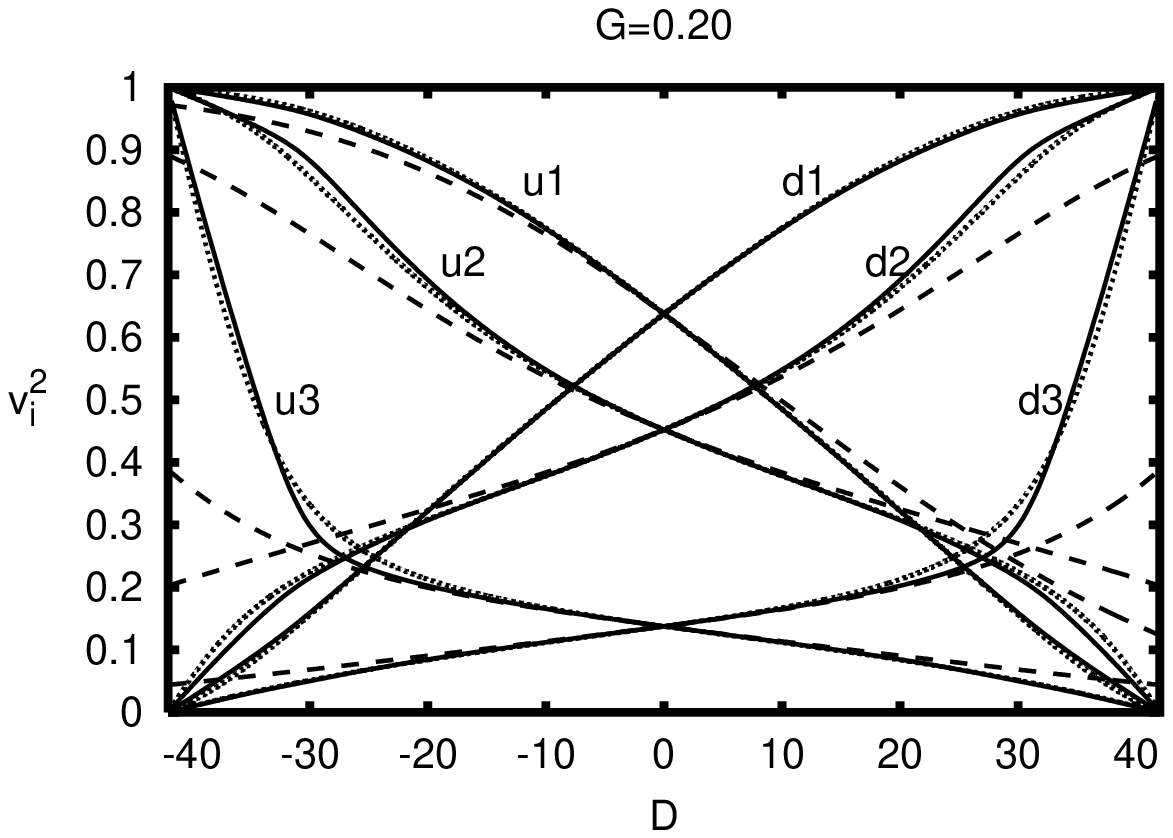}}
\end{tabular}
\end{center}
\caption{
Occupation probabilities $v_i^2$ calculated with the ASCC method 
for $G=0.14$ (top), $G=0.16$ (middle), and $G=0.20$ (bottom)
are plotted by solid lines as functions of the deformation $D$. 
Other parameters used are the same as in Fig.~2. 
For comparison, $v_i^2$ calculated with the CHB method
and with the BCS approximation are also shown by dotted and dashed lines,
respectively. The solid and dotted lines appear similar to each other.
Note that, at the large deformation limit $D_{\rm max}$, 
all down-sloping (up-sloping) levels are fully occupied (unoccupied)
in the cases of the ASCC and CHB methods, whereas it is not the case
in the BCS approximation. }
\label{fig4}
\end{figure}

\normalsize
In Figs. 3 and 4, we have also shown the BCS pairing gaps and occupation 
probabilities
calculated in the BCS approximation. In this approximation, single particle energies given by $e_i=e_i^0-\chi d_i \sigma_i D$
are used to solve the BCS equations, where
the deformation $D$ is treated as an {\it ad hoc} parameter,
neglecting the HB selfconsistency (\ref{defo})
in accord with phenomenological singe-particle potential models. 
The pairing gap $\Delta$ in the BCS approximation differs from those
in the CHB and ASCC methods.
Note that the BCS gaps do not vanish in the limit $D_{\rm max}$
in contrast to the HB selfconsistent pairing gaps.
In the following we shall focus our attention on dynamical 
properties of tunneling motions between the oblate and prolate minima.

\subsection{Excitation spectra and collective mass}

Figure 5 shows the excitation spectra and transition matrix elements
for the ASCC method and for the exact diagonalization in a basis set,
Eq.~(\ref{base}).
Wave functions of the low-lying states are displayed in Figs.~6 and 7.
The collectivity of these states are apparent from the
enhancement of the transition matrix elements
in comparison with the single-particle ones $d_j($=1 or 2).
We see that the ASCC method well reproduces the major characteristics
of the exact spectra and transition properties. 
In view of the huge degrees of freedom involved in the model 
under consideration (the dimension of this shell model space is 1894), 
it is remarkable that the low-lying states can be
described very well in terms of only the single collective coordinate $q$.
In particular, we note that the emergence of
the ``parity splitting'' pattern for weaker 
strength $G$ of the pairing interaction is well reproduced in the
calculation. This implies that the ASCC method succeeds in describing
the large amplitude tunneling motions through the barrier
between the oblate and prolate local minima. 

As is well known, the collective mass parameter represents inertia 
against change of the mean field. It is determined locally and changes 
as $D$ varies. 
Since we have set the scale of the collective coordinate $q$
such that the collective mass parameter $M(q)=B^{-1}(q)$ with respect to $q$
is unity, the collective kinetic energy can be written
either by $q$ or $D$ as
\begin{equation}
\frac{1}{2}\dot{q}^2 = 
\frac{1}{2}(\frac{d q}{d D}\dot{D})^2 
\equiv \frac{1}{2}M(D)\dot{D}^2.
\end{equation}
Thus, we obtain the explicit expression for $M(D)$: 
\begin{equation}
M(D) \equiv (\frac{dq}{dD})^{2}
= (\frac{d}{dq}\sum_{i}2d_{i}\sigma_{i}v_{i}^{2}(q))^{-2}
= (4 \sum_{i} d_{i}\sigma_{i}u_{i}v_{i}P_{i}(q))^{-2}.
\label{massD}
\end{equation}
\noindent 
The collective mass $M(D)$ evaluated in this way is shown in Fig.~8
as a function of $D$. One immediately notice that
the ASCC mass $M(D)$ diverges in the limit $D \to D_{\rm max}$. 
The reason is clearly understood by examining the relationship between 
$D$ and the collective coordinate $q$.
It is shown in Fig.~9. We see that the value of $D$ saturates
as it approach its limiting value $D_{\rm max}$. Thus, the derivative 
$dq/dD$, which corresponds to the collective mass $M(D)$ 
according to Eq.~(\ref{massD}), diverges.
Needless to say, this divergence is caused 
by the existence of the maximum value of $D$ 
which is an artifact of the model: If we increase the number of shells
explicitly taken into account, this limit is removed by successive level 
crosings with increasing $D$.
\begin{figure}[hhhh]
\begin{center}
\begin{tabular}{c}
{\includegraphics[width=7.5cm]{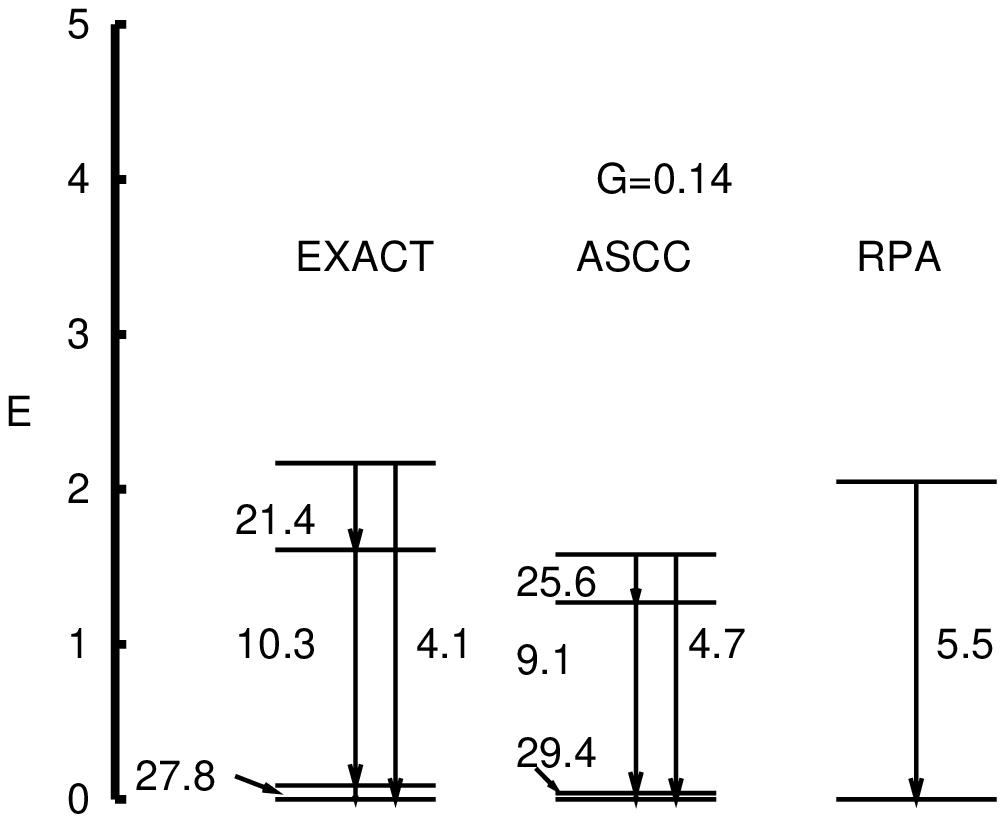}}\\
{\includegraphics[width=7.5cm]{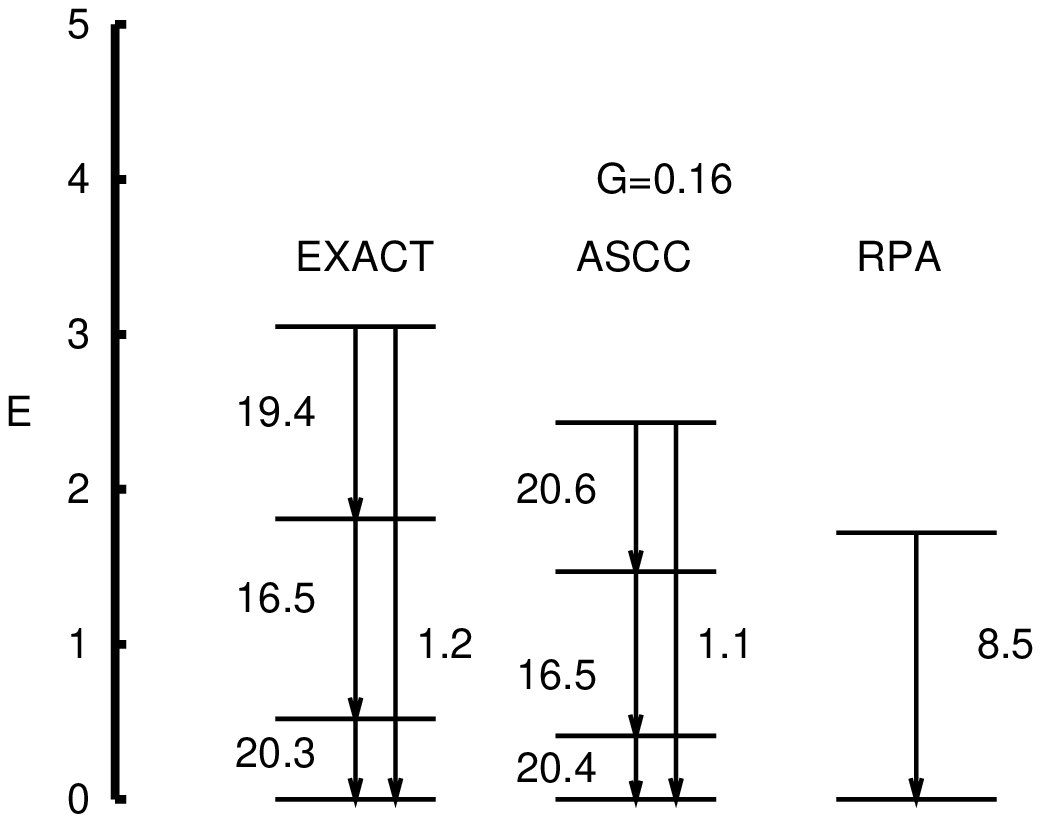}}\\
{\includegraphics[width=7.5cm]{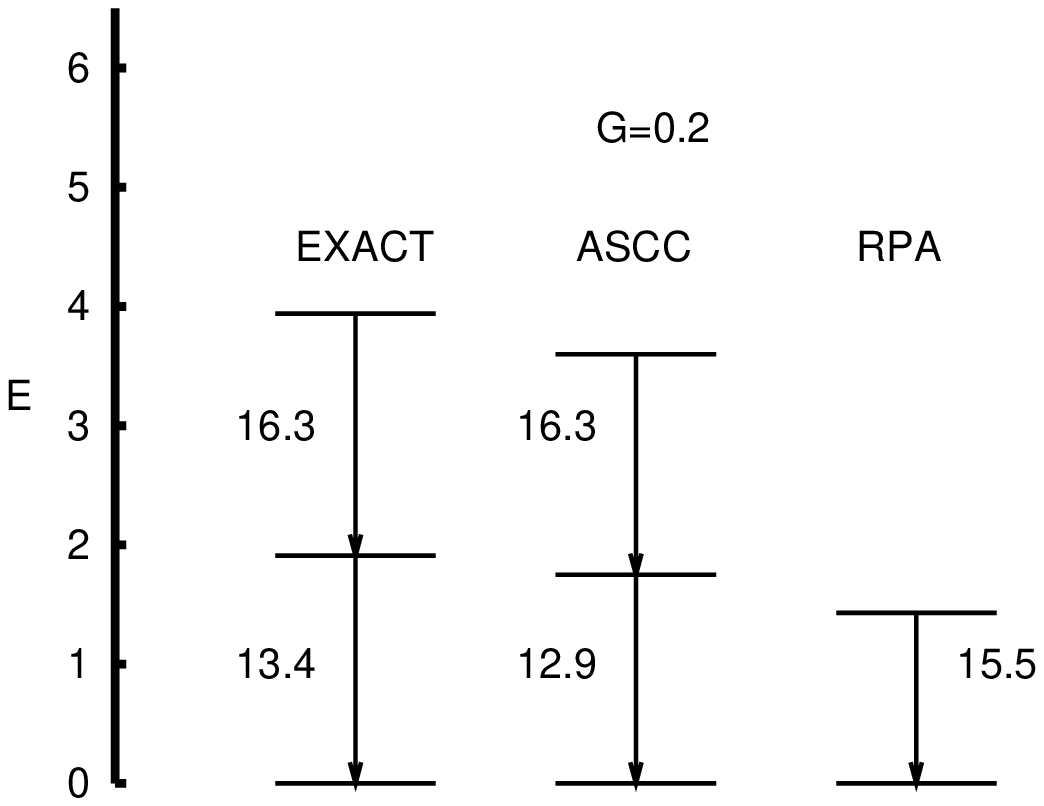}}
\end{tabular}
\end{center}
\caption{
\footnotesize
Comparison of excitation spectra calculated with the ASCC method
and and with the exact diagonalization 
for $G=0.14$ (top), $G=0.16$ (middle), and $G=0.20$ (bottom).
For reference sake, excitation energies of the lowest RPA modes
in the HFB local minima are also indicated.
Other parameters used are the same as in Fig.~2. 
Numbers adjacent to vertical lines indicate
transition matrix elements for the ``quadrupole'' operator ${\hat D}$.
In the top panel, transition matrix elements between
``parity doublets'' are indicated with arrows.}
\label{fig5}
\end{figure}

\begin{figure}[hhhh]
\begin{center}
\begin{tabular}{c}
{\includegraphics[width=7.5cm]{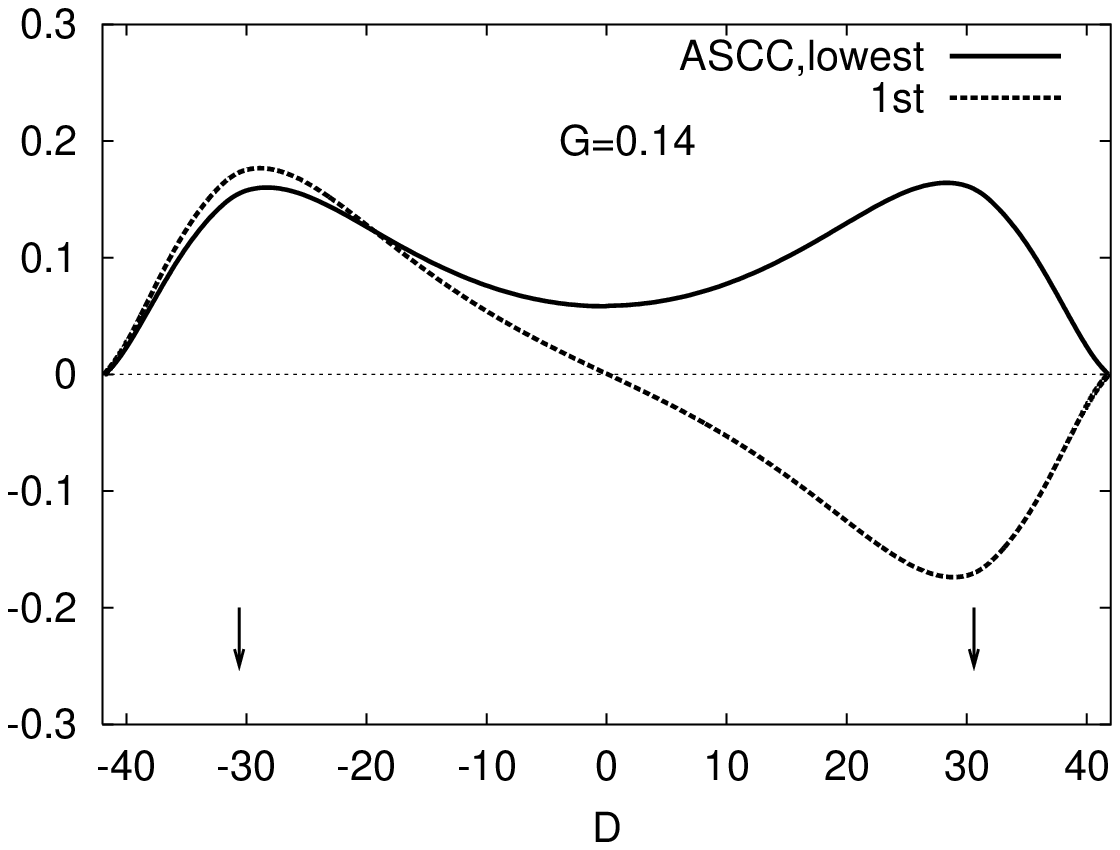}}\\
{\includegraphics[width=7.5cm]{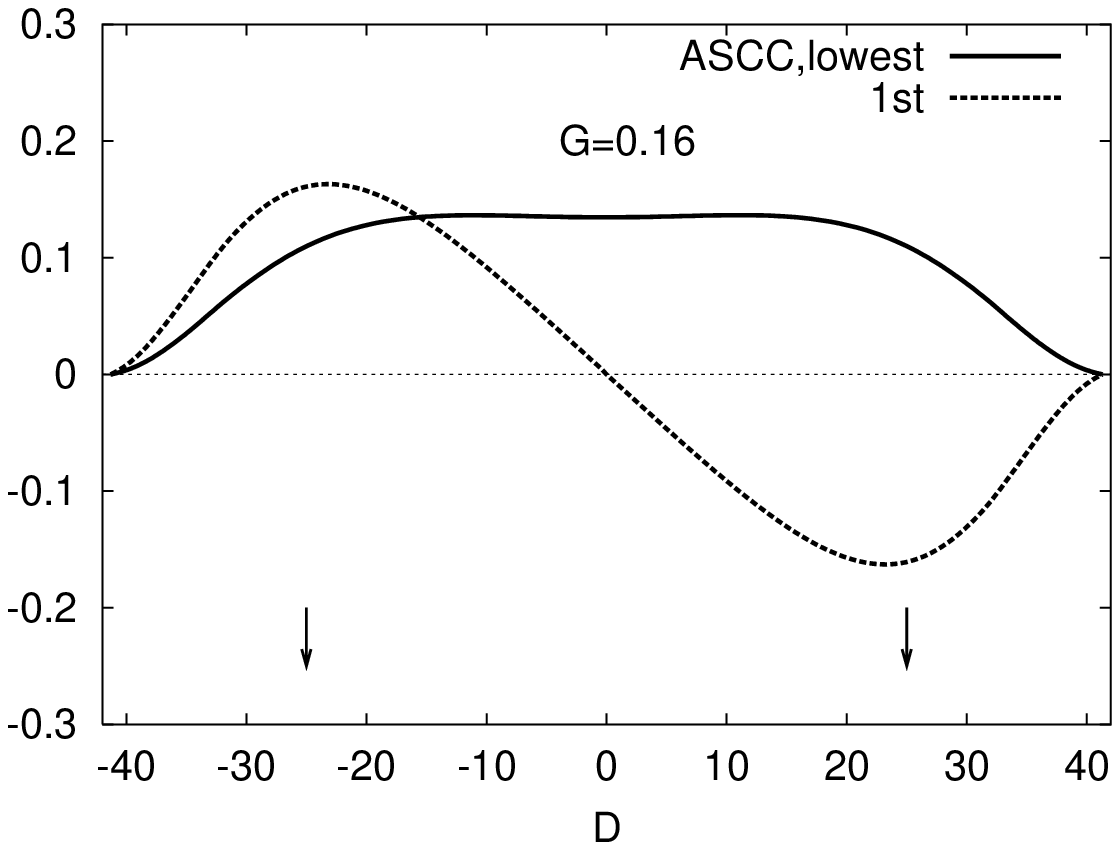}}\\
{\includegraphics[width=7.5cm]{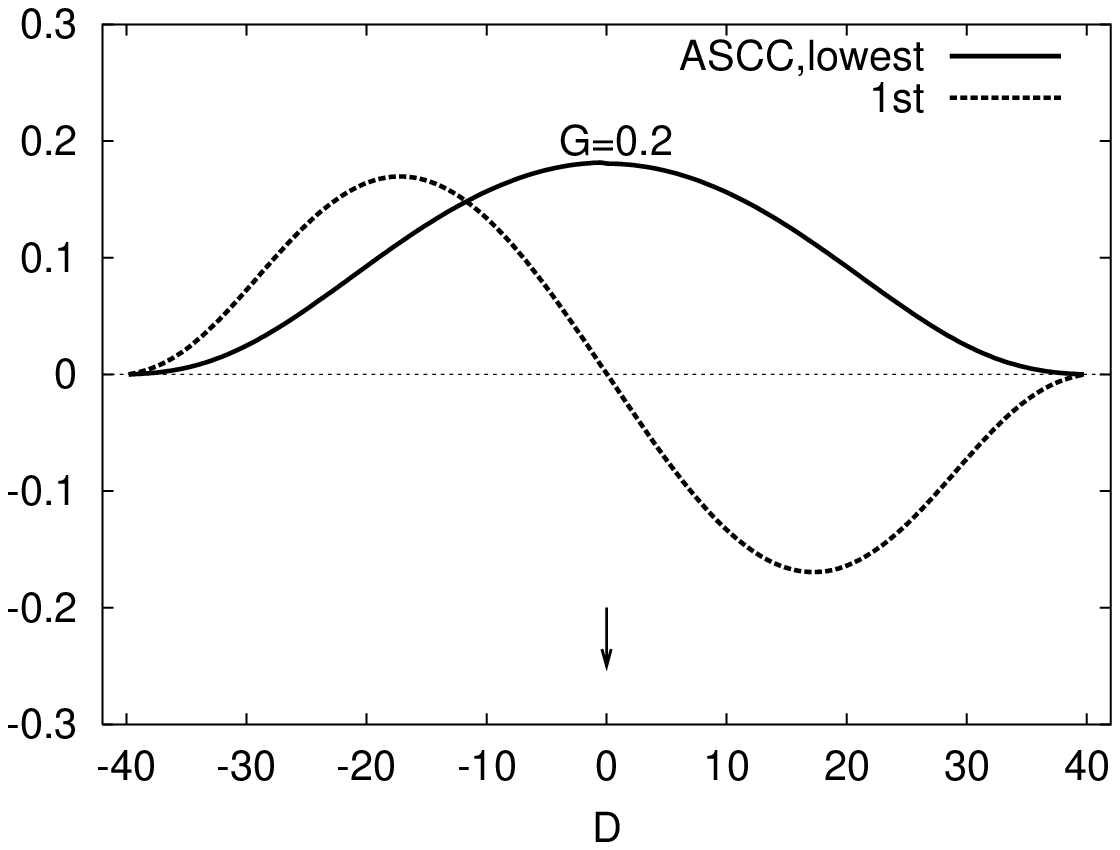}}
\end{tabular}
\end{center}
\caption{
Collective wave functions $\psi(D)$ for the ground and first excited
states in Fig.~5, obtained with the ASCC method for $G=0.14$ (top), $G=0.16$ (middle), and $G=0.20$ (bottom). 
The equilibrium deformations are indicated by arrows.
The $\psi(D)$ are defined by 
$\psi(D)=\psi(q)|\frac{\partial q}{\partial D}|^{\frac{1}{2}}$,
where $\psi(q)$ are wave functions for the collective coordinate $q$. 
Thus, they are normalized as
$\int |\psi(q)|^2dq=\int |\psi(q)|^2|\frac{\partial q}{\partial D}|dD=
\int |\psi(D)|^2dD=1$.}
\label{fig6}
\end{figure}

\begin{figure}[hhhh]
\begin{center}
\begin{tabular}{c}
{\includegraphics[width=7.5cm]{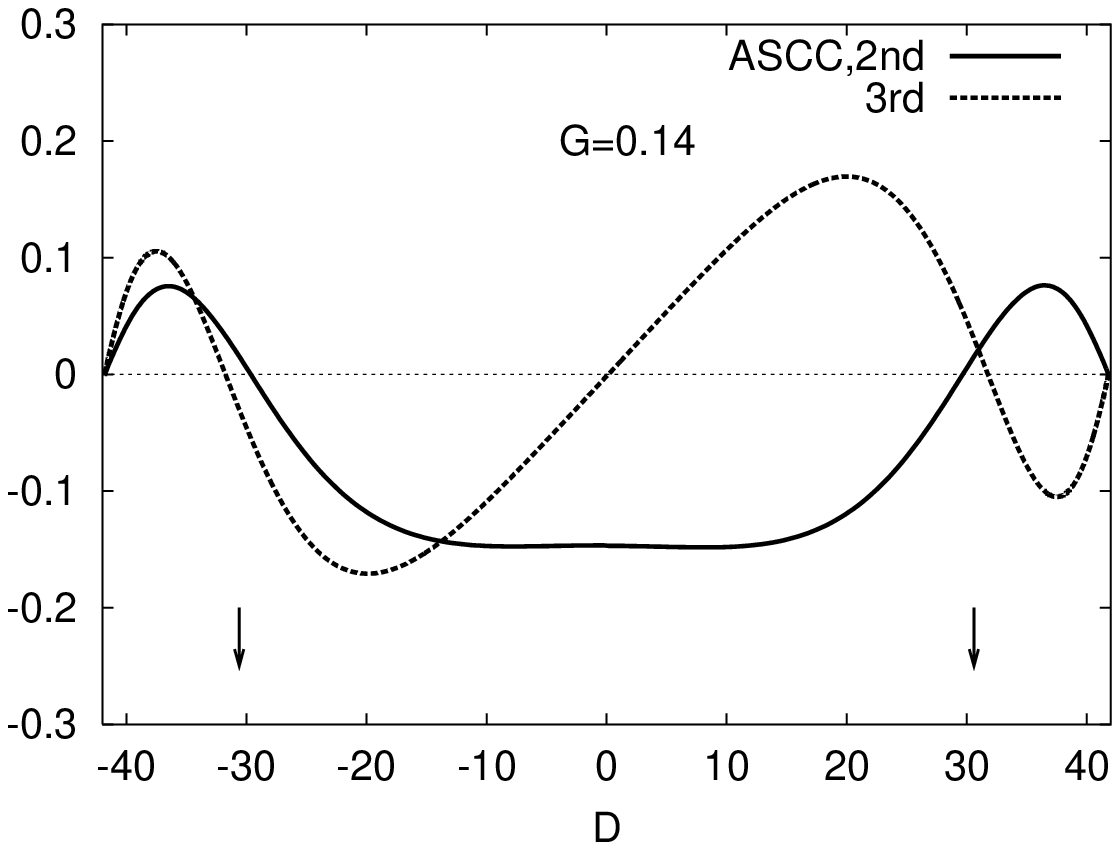}}\\
{\includegraphics[width=7.5cm]{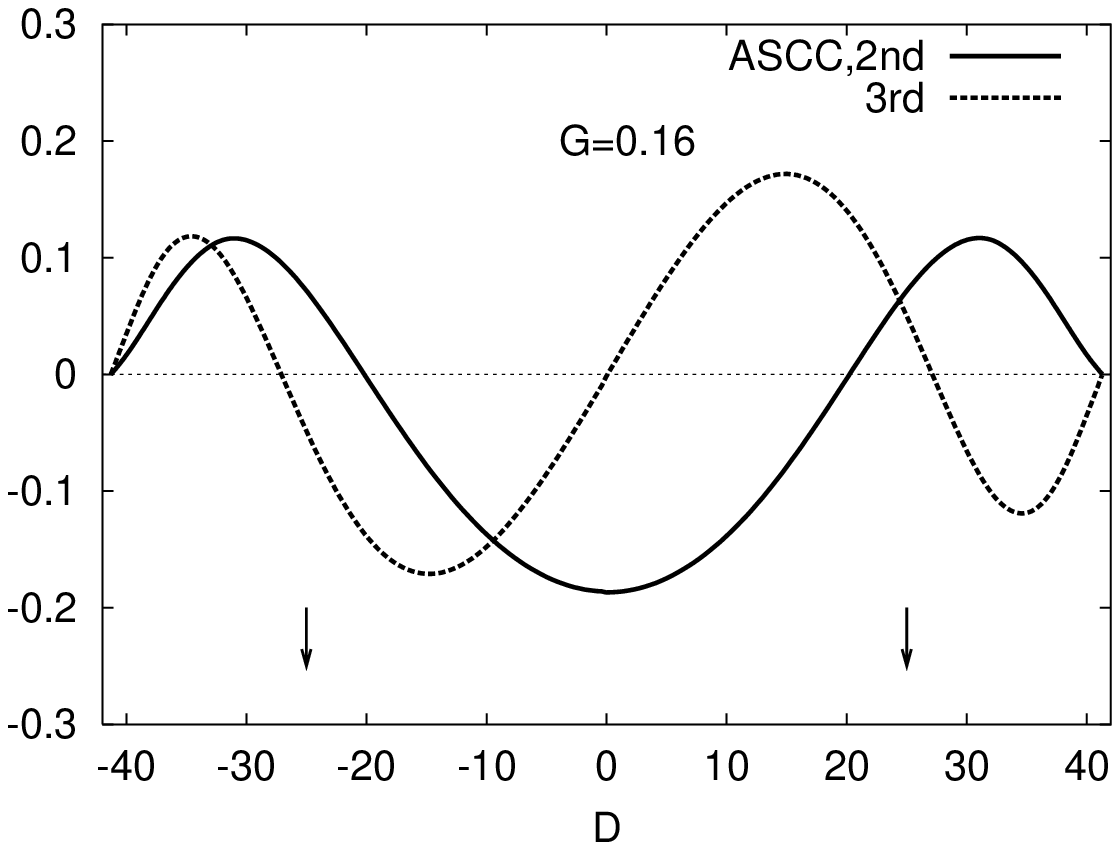}}\\
{\includegraphics[width=7.5cm]{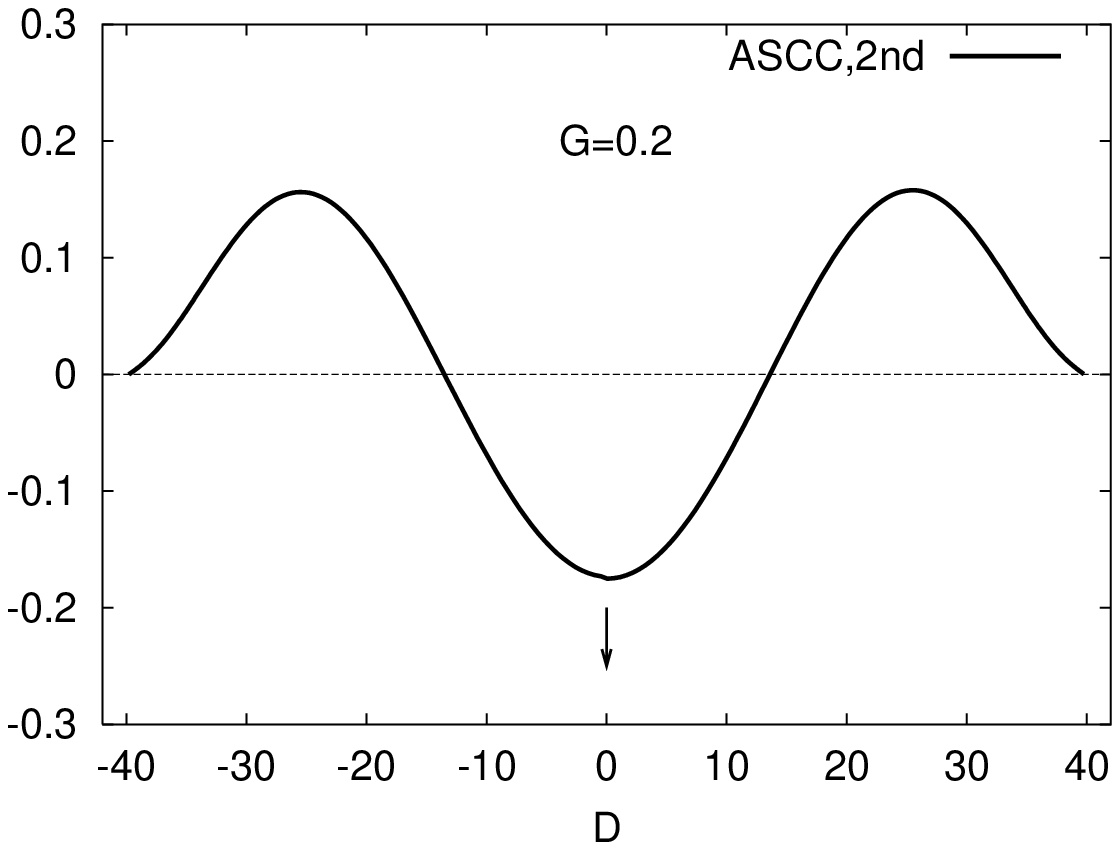}}
\end{tabular}
\end{center}
\caption{\small
The same as Fig.~6 but for the second and third excited states.
For $G=0.2$, only the wave function for the second excited state is drawn.}
\label{fig7}
\end{figure}

\begin{figure}[hhhh]
\begin{center}
\begin{tabular}{c}
{\includegraphics[width=7.5cm]{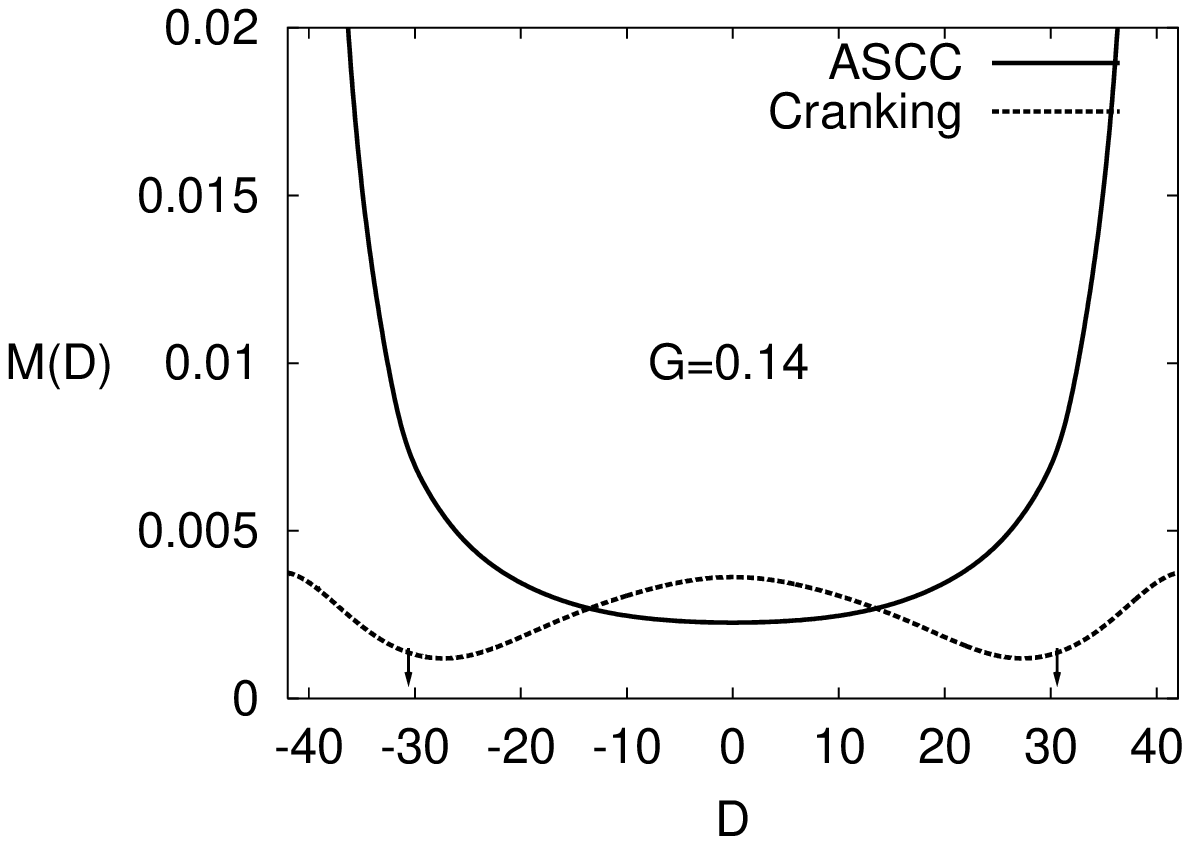}}\\
{\includegraphics[width=7.5cm]{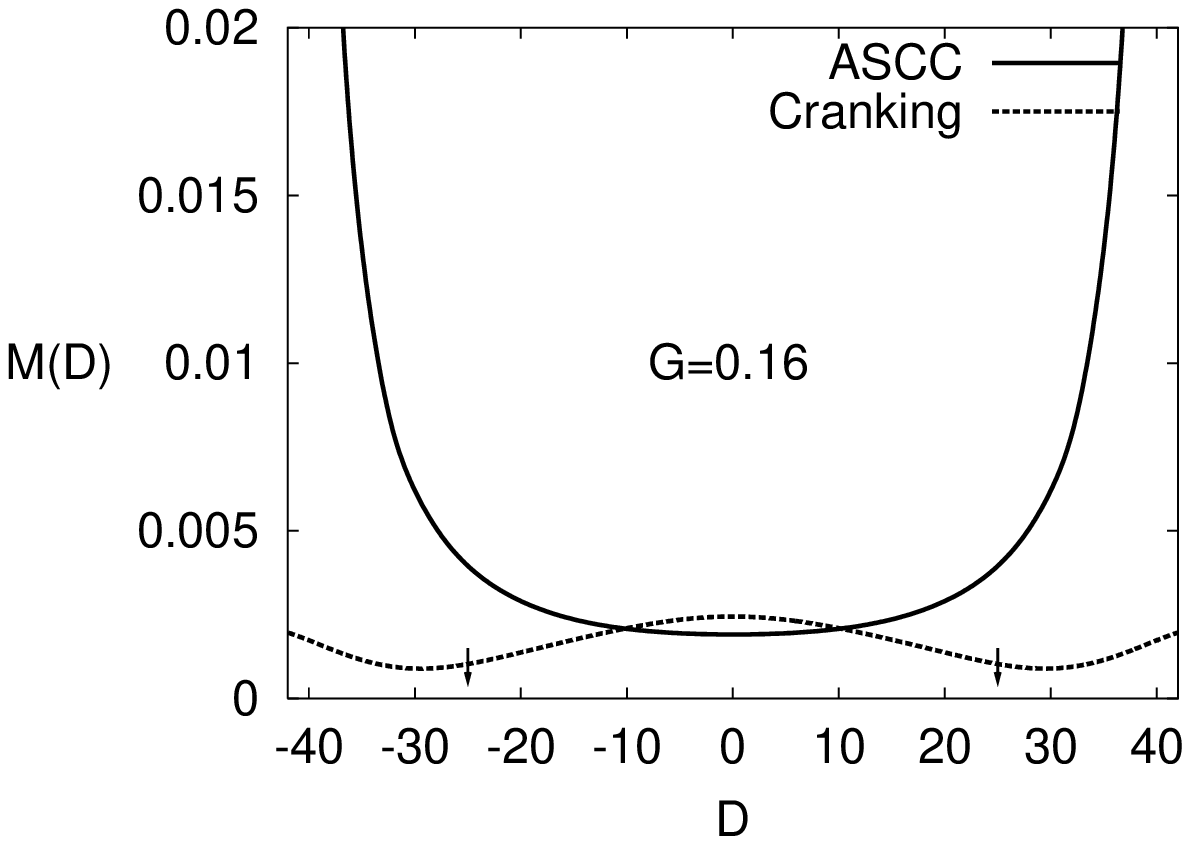}}\\
{\includegraphics[width=7.5cm]{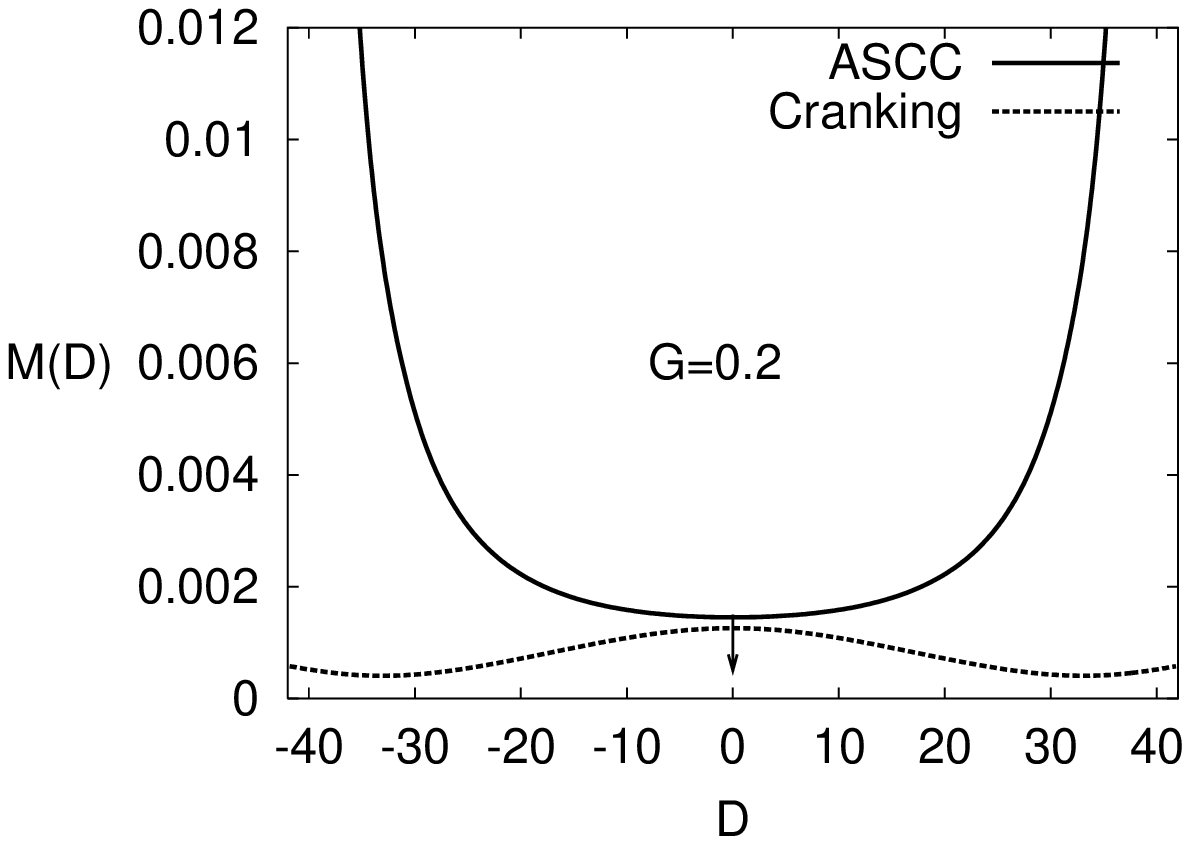}}
\end{tabular}
\end{center}
\caption{ \footnotesize
Collective mass $M(D)$ calculated with the ASCC method 
for $G=0.14$ (top), $G=0.16$ (middle), and $G=0.20$ (bottom)
are plotted by solid lines as functions of deformation $D$. 
Other parameters used are the same as in Fig.~2. 
For comparison, the cranking mass are indicated by dotted lines.
The equilibrium deformations are indicated by arrows.}
\label{fig8}
\end{figure}

\begin{figure}[hhhh]
\begin{center}
{\includegraphics[width=7.5cm]{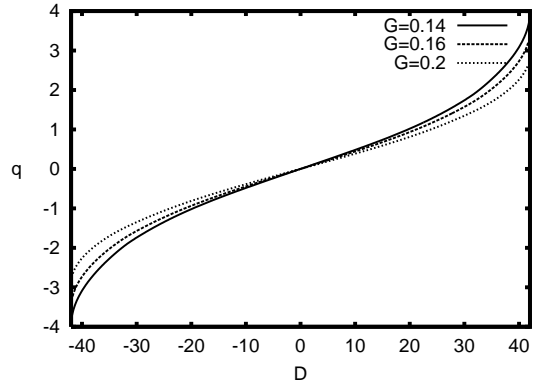}}
\end{center}
\caption{\small{
Relations between the deformation $D$ and the
collective coordinate $q$ in the ASCC method.
The solid, dashed and dotted lines show the results of
calculation for $G=0.14, 0.16$, and 0.20, respectively.}}
\label{fig9}
\end{figure}

\normalsize

It is interesting to compare the ASCC mass $M(D)$ with
the cranking mass $M_{\rm cr}(D)$\cite{rf:1}, 
\begin{eqnarray}
M_{\rm cr}(D)
&=&2\sum_n 
\frac{ | \langle\phi_n^{\rm BCS}(D)|\frac{\partial}{\partial D}
|\phi_0^{\rm BCS}(D)\rangle |^2}
{E_n(D) - E_0(D)}\nonumber \\
&=&2\sum_{i} 
\frac{ |
2u_i v_i(\chi d_{i}\sigma_{i} + \frac{\partial \lambda}{\partial D})
+ (u^2_i-v^2_i)\frac{\partial \Delta}{\partial D} |^2 }
{(2E_i)^3}
\label{massqc}
\end{eqnarray}
\noindent
where $\phi_0^{\rm BCS}(D)$ and $\phi_n^{\rm BCS}(D)$ represent the ground 
and excited states, 
and $E_0(D)$ and $E_n(D)$ the energies of them 
obtained in the BCS approximation.
Namely, the coefficients of the Bogoliubov transformations,
$u_i, v_i$, the pairing gap $\Delta$, the chemical potential $\lambda$,
and the quasiparticle energies $E_i$ are evaluated with use of 
single-particle energies defined by $e_i=e_i^0-\chi d_i \sigma_i D$.
It is important to note that the deformation $D$ is treated 
in the BCS approximation 
as an {\it ad hoc} potential parameter disregarding 
the selfconsistency condition (\ref{defo}).
Figure~8 shows that the cranking mass $M_{\rm cr}(D)$ is significantly
different from the ASCC mass $M(D)$ in the whole interval of
the deformation $D$, including the spherical point $D=0$ and the
deformed equilibrium points. The difference between the ASCC mass
and the cranking mass can be understood in terms of the HB selfconsistency. 

At the spherical point ($D=0$), 
we can make the comparison between the ASCC mass and the cranking mass 
in an explicit way. 
There, all terms linear with respect to
$\sigma_i$ in the local harmonic equations 
vanish after summing over all levels $i$ so that  
the pairing and ``quadrupole'' normal modes are exactly decoupled.
Thus, we get a simple expression of the ASCC mass
\begin{equation}
M(D=0)=2\chi^2\sum_i
\frac{2E_i(2d_i \sigma_i u_i v_i)^2}{((2E_i)^2-\omega^2)^{2}}
= 4 \chi^2\Delta^2\sum_i \frac{d_i^2}{E_i ((2E_i)^2-\omega^2)^2}.
\label{massq0}
\end{equation}
\noindent
On the other hand, 
the expression of the cranking mass $M_{\rm cr}(D)$ reduces to
\begin{equation}
M_{\rm cr}(D=0)
=\frac{1}{4}\chi^{2}\Delta^{2}\sum_{i}\frac{d_{i}^{2}}{E_{i}^{5}},
\label{massqc}
\end{equation}
since the derivatives, $\partial\Delta/\partial D$ and
$\partial\lambda/\partial D$ vanish at $D=0$. 
\noindent
We see that the above expression for $M_{\rm cr}(D=0)$ 
can be related to $M(D=0)$ if we set $\omega=0$ in Eq.~(\ref{massq0}).
In reality, the frequency $\omega$ is imaginary and $\omega^2$ is negative
when the barrier exists ($G=0.14$ and 0.16). Accordingly, 
$M(D=0) \leq M_{\rm cr}(D=0)$
in these cases. On the other hand, $\omega$ is real and $\omega^2$ is positive
when the spherical point is stable ($G=0.20$), 
so that $M(D=0) \geq M_{\rm cr}(D=0)$ in this case.
In this way, the difference between the ASCC mass and the cranking mass 
in the barrier
region (near $D \approx 0$) noticeable in Fig.~8 can be understood in terms
of the finite frequency effect of the local harmonic mode, 
which is connected with the curvature of the collective potential 
by Eq.~(\ref{freq}).  
This finite frequency effect decreases (increases) the collective mass  
when the barrier emerges (diminishes).
We would like to emphasize that 
the finite frequency effect in the local harmonic equations 
represents selfconsistent dynamics of the time-dependent mean-field.

We note that, in contrast to the ASCC mass $M(D)$,
the cranking mass $M_{\rm cr}(D)$ does not diverge at $D_{\rm max}$.
This is because, as already emphasized above, the deformation $D$ is 
treated as an {\it ad hoc} parameter in the single-particle potential, 
so that the existence of the limit $D_{\rm max}$ in the selfconsistent 
deformation defined by Eq.~(\ref{defo}) is disregarded there. 
In the multi-$O(4)$ model under consideration, the selfconsistency between
the deformation parameter specifying the single-particle potential 
and the density deformation evaluated in terms of the wave function
becomes extremely important near $D_{\rm max}$.

The comparison between the ASCC and cranking masses in Fig.~8
thus indicates the importance of the HB selfconsistency 
in evaluating the collective mass, although it is
somewhat exaggerated because of an artificial aspect
of the present parameter setting of the multi-$O(4)$ model.

There are various microscopic approaches to derive the collective
mass (also called inertial functions)\cite{rf:1, rf:2,rf:31}.
Certainly, it is important and interesting to make a detailed comparison of 
different approaches for the collective mass in the multi-$O(4)$ model,
and clarify the role of the HB selfconsistency in more detail.
Such a more systematic and comparative study is beyond the scope of this paper, but will be pursued in a separate paper.  

\section{Conclusions}

The ASCC method was applied to
the exactly solvable multi-$O(4)$ model which simulates nuclear shape 
coexistence phenomena. The collective mass and
dynamics of large amplitude collective motions in this model system
were analysed, and it was shown that the method can well describe
the tunneling motions through the barrier between
the prolate and oblate local minima in the collective potential.
The result of numerical analysis strongly encourages
applications of this approach to realistic cases. We plan to investigate
the oblate-prolate shape coexistence phenomena in $^{68}$Se with use of
the P+Q interactions.


%

\end{document}